\documentclass[11pt,a4paper,onepage]{amsart}
\usepackage{eurosym}
\usepackage{enumerate}

\usepackage[a4paper,twocolumn=false,twoside=false,
 total={176mm,257mm},
 left=17mm,
 top=20mm]{geometry}

\usepackage[pdftex]{graphicx}
\usepackage{amsmath,amsfonts,amssymb}
\usepackage{dsfont}
\usepackage{algorithmic}
\usepackage{array}
\usepackage[caption=false,font=footnotesize]{subfig}
\usepackage{stfloats}
\usepackage{bigints}
\usepackage{url}
\usepackage{color}
\newcommand{\sumweyl}{\sum_{w\in W}}

 \usepackage[linktocpage=true,colorlinks=true,citecolor=blue,linkcolor=blue,urlcolor=blue]{hyperref}

\usepackage[numbers,compress,square,comma]{natbib}

\DeclareMathOperator{\R}{\mathbb{R}}

\begin{document}
\bibliographystyle{ourstyle}

\title[]{Riemannian Gaussian distributions, random matrix ensembles and diffusion kernels}

\author[]{Leonardo Santilli${}^{\ast}$}
\address[$\ast$]{Departamento de Matem\'{a}tica, Grupo de F\'{i}sica Matem\'{a}tica, Faculdade de Ci\^{e}ncias, Universidade de Lisboa, Campo Grande, Edif\'{i}cio C6, 1749-016 Lisboa, Portugal.}
\email{lsantilli@fc.ul.pt}
\author[]{Miguel Tierz${}^{\dagger,\ddagger}$}
\address[$\dagger$]{Departamento de Matem\'{a}tica, ISCTE - Instituto Universit\'{a}rio
de Lisboa, Avenida das For\c{c}as Armadas, 1649-026 Lisboa, Portugal.}
\email{mtpaz@iscte-iul.pt}
\address[$\ddagger$]{Departamento de Matem\'{a}tica, Grupo de F\'{i}sica Matem\'{a}tica, Faculdade de Ci\^{e}ncias, Universidade de Lisboa, Campo Grande, Edif\'{i}cio C6, 1749-016 Lisboa, Portugal.}
\email{tierz@fc.ul.pt}

\maketitle

\begin{abstract}
\vspace{-0.5cm}
We show that the Riemannian Gaussian distributions on symmetric spaces, introduced in recent years, are of standard random matrix type. 
We exploit this to compute analytically marginals of the probability density functions. This can be done fully, using Stieltjes-Wigert orthogonal polynomials, for the case of the space of Hermitian matrices, where the distributions have already appeared in the physics literature. For the case when the symmetric space is the space of $m \times m$ symmetric positive definite matrices, we show how to efficiently compute densities of eigenvalues by evaluating Pfaffians at specific values of $m$. Equivalently, we can obtain the same result by constructing specific skew orthogonal polynomials with regards to the log-normal weight function (skew Stieltjes-Wigert polynomials). Other symmetric spaces are studied and the same type of result is obtained for the quaternionic case. Moreover, we show how the probability density functions are a particular case of diffusion reproducing kernels of the Karlin-McGregor type, describing non-intersecting Brownian motions, which are also diffusion processes in the Weyl chamber of Lie groups.
\end{abstract}

\section{Introduction}

The statistical analysis of a probability measure $Q$ on a differentiable manifold is a topic attracting much attention and has developed considerably due to the wide range of interesting applications that require the analysis of data not living in ordinary Euclidean spaces. There is great interest, for example, in trying to process datasets which lie in the space $\mathcal{P}_{m\,}$ of $m \times m$ symmetric positive definite matrices.

This space can be equipped with a Riemannian metric, giving it the structure of a Riemannian homogeneous space of negative curvature. Such Riemannian manifold has been thoroughly studied from many points of view, corresponding to different disciplines, such as harmonic analysis and number theory~\cite{helgason,terras2}, matrix analysis~\cite{bhatia} and information geometry and multivariate statistics \cite{statistics,atkinson}. 

This and other symmetric spaces have been studied endowed with metrics such as the Rao-Fisher metric or the log-Euclidean metric~\cite{pennec3}, in order to develop the tools to properly carry out statistical inference of data on such spaces. The landscape of applications is very vast as it includes, for example, medical imaging~\cite{pennec2,congedo}, continuum mechanics~\cite{moakher2}, radar signal processing~\cite{arnaudon,arnaudon1} and computer vision~\cite{image1,image2,dong,tuzel,caseiro}, among many other references that could be quoted here.

In the present paper, the Riemannian metric will be the Rao-Fisher metric, also known as affine-invariant metric, and is the subject of Section \ref{subsec:metric}. A specific probabilistic model to represent the statistical variability of data in non-Euclidean spaces, such as $\mathcal{P}_{m}$, has been developed by introducing the so-called Gaussian probability distributions on Riemannian symmetric spaces \cite{cheng2013novel,said2017riemannian,said2017gaussian,said2019warped}.

The Riemannian Gaussian distribution $G(\bar{Y},\sigma)$ will depend on two parameters, $\bar{Y} \in \mathcal{P}_m$ and $\sigma > 0$. The form of its probability density function, generalising that of a Gaussian distribution on the Euclidean space $\mathbb{R}^{p}$, is given by
\begin{equation} \label{eq:gaussianpdf}
p(Y | \,\bar{Y},\sigma) = \frac{1}{\zeta(\sigma)} \, \exp \left[ - \frac{d^{\,2}(Y,\bar{Y})}{2\sigma^2}\,\right],
\end{equation} 
where $d : \mathcal{P}_m \times \mathcal{P}_m \rightarrow \mathbb{R}_+$ is Rao's Riemannian distance (defined below), and the density is with respect to the Riemannian volume element of $\mathcal{P}_{m\,}$, denoted by $dv(Y)$. In comparison to a Gaussian distribution on $\mathbb{R}$, the normalising factor $\zeta(\sigma)$ plays the role of the factor $\sqrt{\,2\pi\,\sigma^{\scriptscriptstyle 2}}\,$.

Among other possible applications, the distributions provide a statistical foundation for the concept of Riemannian centre of mass, which plays a major role in many contexts~\cite{afsari,moakher1}. They also play a role in machine learning when the dataset is inherently structured \cite{mathieu2019continuous,ovinnikov2020poincare,gerald2020node}.

In addition to the relationship between Riemannian Gaussian distributions and the concept of Riemannian centre of mass and the potential for applications, discussed in \cite{cheng2013novel,said2017riemannian,said2017gaussian,said2019warped}, we shall show here that one of the hallmarks of these distributions is analytical tractability. In particular, we will show how the obtained distributions are of standard random matrix type and how the ensuing analytical tools associated to the study of random matrices can be used effectively to further characterize the distributions analytically \cite{MehtaBook,ForresterBook,livan2018introduction}. 

More specifically, we shall use Pfaffian and determinant descriptions of the random matrix ensembles and the classical orthogonal polynomial method in random matrix theory \cite{MehtaBook,ForresterBook,livan2018introduction}. We do this in the setting which includes skew-orthogonal polynomials, necessary to solve models which do not correspond to the case of Hermitian matrices, such as the case of the real symmetric matrices, which is central for statistical applications.

We now briefly summarize the derivation of the model for the case of real symmetric matrices, along the lines of \cite{cheng2013novel,said2017riemannian,said2017gaussian,said2019warped}, to set up the context. Further details are in these references, while we will focus, from Section \ref{sec:MM} on, on the analytical characterization of the joint probability distribution functions using random matrix theory.

\subsection{The Rao-Fisher metric: distance, geodesics and volume}
\label{subsec:metric}
We overview the general setup, following closely the presentation in \cite{said2017riemannian}.\par
The Rao-Fisher metric \cite{atkinson,terras2} 
\begin{equation} \label{eq:metric}
   ds^2(Y) = \mathrm{tr}\, [Y^{-1}dY]^2 , \qquad Y \in \mathcal{P}_m 
\end{equation}
is a Riemannian metric on $ \mathcal{P}_m$. It induces a Riemannian distance $d : \mathcal{P}_m \times \mathcal{P}_m \rightarrow \mathbb{R}_+$, known as Rao's distance, in the standard way.

When equipped with the Rao-Fisher metric, the space $\mathcal{P}_m$ becomes a Riemannian manifold of negative sectional curvature~\cite{helgason,terras2}. Since it is also complete and simply connected, the distance between any two points in $\mathcal{P}_m$ is given by the length of the geodesic connecting them.

The volume form on $\mathcal{P}_m$ induced by the Rao-Fisher metric is~\cite{terras2} 
\begin{equation} \label{eq:dv}
  dv(Y) = \det(Y)^{-\frac{m+1}{2}}\prod_{i \leq j} dY_{ij}
\end{equation}
where indices denote matrix entries. 

One can then consider the expressions of $ds^2(Y)$ and $dv(Y)$, given by \eqref{eq:metric} and \eqref{eq:dv}, in terms of polar coordinates \cite{ForresterBook}. Then, for any \emph{class function} $f : \mathcal{P}_m \rightarrow \mathbb{R}$, the angular coordinates can be integrated out, obtaining
\begin{equation} \label{eq:integrationew}
  \int_{\mathcal{P}_m} f(Y) \, dv(Y) =  \frac {8^{\frac{m(m-1)}{4}} }{m! \, 2^m } \, \omega_{m} \int_{\mathbb{R}^m}  f(r_1, \dots, r_m) 
\prod_{i < j }  \sinh\left(\frac{|r_i - r_j|}{2}\right) \prod^m_{i=1} dr_i, 
\end{equation}
where (see \cite[Page 71]{muirhead}), 
\begin{equation}
  \omega_m  \equiv \frac{2^m \pi^{m^2/2}}{\strut \Gamma_m (m/2)}
\label{eq:omegam}
\end{equation}
with $\Gamma_m$ the multivariate Gamma function, given in~\cite{muirhead}. Formula \eqref{eq:integrationew} follows directly from~\cite[Proposition 3, Page 43]{terras2} but has been written following the notation in \cite{said2017gaussian}.\par
\medskip

For any $\sigma > 0$, let $f(Y \vert \sigma)$ be given by,
\begin{equation} \label{eq:f}
f(Y \vert \sigma ) = \exp \left[ - \frac{d^{\,2}(Y,I)}{2\sigma^2}\,\right] 
\end{equation}
where $I \in \mathcal{P}_{m\,}$ is the $m \times m$ identity matrix. The normalization constant $\zeta(\sigma)$ is then defined as 
\begin{equation} \label{eq:prez}
\zeta(\sigma) = \int_{\mathcal{P}_m} f(Y \vert \sigma  ) \, dv(Y) ,
\end{equation}
where $dv(Y)$ is the volume form \eqref{eq:dv}. Then, 
\begin{equation} \label{eq:z}
\zeta(\sigma) = \omega_m \, \cdot \, \frac{ 8^{\frac{m(m-1)}{4}} }{ m!\, 2^m } \,  \int_{\mathbb{R}^m} e^{- (r^{\,2}_1 + \cdots + r^{\,2}_m)/2\sigma^2}  \, \prod_{1 \le i < j \le m } \sinh\left(|r_i - r_j|/2\right) \, \prod^m_{i =1 } dr_i
\end{equation}
with $\omega_m$ defined in \eqref{eq:omegam}.\par
\medskip
We show how to use random matrix tools, in particular the evaluation of Pfaffians, to compute analytically $\zeta(\sigma)$ for several values of $m$ and also in the large $m$ limit. This extends some of the results in \cite{cheng2013novel,said2017riemannian} and the other works discussing Gaussian distributions on Riemannian symmetric spaces, where the case $m=2$ is evaluated analytically and the rest is carried out numerically, with Monte-Carlo methods. Before that, we note that the corresponding model for Hermitian matrices, also discussed in \cite{said2017gaussian}, which is characterized by an analogous expression but with the $\sinh\left(|r_i - r_j|/2\right)$ term squared, is fully solvable with orthogonal polynomials. It has already been studied in the physics literature in detail, but we summarize and develop the results in the next Section and we will study more difficult cases to treat analytically, such as \eqref{eq:z}, in Section \ref{sec:beta-14} and in the Appendix. Finally, Section \ref{sec:diffusion} will contain a discussion of these statistical distributions from the point of view of diffusion processes.

\section{Full solution of the case of Hermitian matrices}
\label{sec:MM}

Let us define the function $z_{\beta} (\sigma)$ as 
\begin{equation}
\label{ZCS}
    z_{\beta} (\sigma) = \frac{1}{m!} \int_{\R^m} \prod_{1 \le i < j \le m} \left[ 2 \sinh \left( \frac{ \vert r_i - r_j \vert }{2} \right) \right]^{\beta} \prod_{i=1} ^{m} e^{- c_{\beta} \frac{r_i^2}{\sigma^2}} \frac{d r_i}{\sqrt{\pi \sigma^2}} ,
\end{equation}
that we call the \emph{partition function}, adopting the language of random matrix theory. The parameter $\beta$ is called Dyson's index.
While \eqref{ZCS} makes sense for every $\beta>0$, with a suitable choice of $c_{\beta}>0$, we will be mainly interested in $\beta \in \left\{1,2,4 \right\}$, corresponding to a random matrix ensemble with orthogonal, unitary and symplectic symmetry, respectively. To avoid cumbersome overall factors of $2^{m/2}$ in what follows, we adopt the normalization of \cite{AFNvM} and choose 
\begin{equation}
\label{normcbeta}
    c_{\beta}= \begin{cases} \frac{1}{2} & \beta=1 , \\ 1 & \beta\in \left\{2,4 \right\}, \end{cases}
\end{equation}
but we stress that other suitable choices exist, as for example $c_{\beta}=\frac{\beta}{2}$, and the present discussion does not depend on such choice. Note that the variance with this normalization is $\frac{ \sigma^2}{2 c_{\beta}}$, so in particular it is $\sigma^2$ for $\beta=1$ but it is $\frac{\sigma^2}{2}$ for $\beta=2$. For $\beta=1$ in \eqref{ZCS} we find
\begin{equation}
\label{eq:ztozeta}
    \zeta (\sigma) = \frac{2^{\frac{m (m-1)}{4}} \pi^{{\frac{m (m+1)}{2}}} }{\Gamma_m (m/2) } \vert \sigma \vert^{m} z_{1} (\sigma) .
\end{equation}\par
The crucial fact we want to exploit is that $z_{\beta} (\sigma)$ can be written in fully standard random matrix form, meaning in terms of a Vandermonde determinant. This was done in \cite{Forrester:Vicious}, in the context of the study of vicious walkers\footnote{We detail further the connection with diffusion in Section \ref{sec:diffusion}.} and, later on, it was also crucial to study Chern-Simons gauge theories \cite{Tierz:02}. One can therefore, use all the power of the traditional random matrix tools, such as determinantal expressions and the method of orthogonal polynomials. Denoting $x_i = e^{u_i }$ and using
\begin{equation}
   \prod_{1 \le i < j \le m} \left\lvert x_i - x_j \right\rvert^{\beta} = \prod_{1 \le i < j \le m} \left\lvert e^{u_i} - e^{u_j} \right\rvert^{\beta} = \left[  \left( \prod_{i=1}^m e^{\frac{u_i}{2} (m-1) }  \right) \prod_{1 \le i < j \le m}  2 \sinh \left( \frac{ \vert u_i - u_j \vert }{2} \right) \right]^{\beta} ,
   \label{change}
\end{equation}
we see that 
\begin{multline}
    \int_{(0, \infty)^m} \prod_{1 \le i < j \le m} \left\lvert x_i - x_j \right\rvert^{\beta}  \prod_{i=1}^{m} e^{- \frac{c_{\beta}}{\sigma^2} \left( \log x_i \right)^2 }  \frac{dx_i}{\sqrt{\pi \sigma^2} } \\ = \int_{\R^m} \prod_{1 \le i < j \le m} \left[  2 \sinh \left( \frac{ \vert u_i - u_j \vert }{2} \right) \right]^{\beta} \prod_{i=1} ^{m} e^{- c_{\beta} \frac{u_i^2}{\sigma^2} + u_i \frac{\beta}{2} \left( m -1 + \frac{2}{\beta} \right) } \frac{d u_i}{\sqrt{\pi \sigma^2}} .
\end{multline}
Changing variables $r_i = u_i - \frac{\sigma^2 \beta}{4 c_{\beta}} \left( m -1 + \frac{2}{\beta} \right)  $ to complete the square, it follows that:
\begin{equation}
\label{ZSW}
    z_{\beta} (\sigma) = \frac{e^{- \sigma^2 \frac{\beta^2}{16 c_{\beta} } m  \left( m -1 + \frac{2}{\beta} \right)^2 } }{m!}   \int_{(0, \infty)^m} \prod_{1 \le i < j \le m} \left\lvert x_i - x_j \right\rvert^{\beta}  \prod_{i=1}^{m} e^{- \frac{c_{\beta}}{\sigma^2} \left( \log x_i \right)^2 }  \frac{dx_i}{\sqrt{\pi \sigma^2} } 
\end{equation}
which describes the partition function of the Stieltjes-Wigert random matrix model. Note that this model is of the standard random matrix form but with a weight function of log-normal type: $w(x)=e^{- \frac{c_{\beta}}{\sigma^2} \left( \log x_i \right)^2 }$. Random matrix theory gives formulas\footnote{For the case $\beta=2$. The other cases are more complicated as they require skew-orthogonal polynomials.} for any marginal of the joint probability density function of eigenvalues, including the normalization constant, in terms of the polynomials orthogonal with regards to the log-normal weight $w (x)$ above. These polynomials are the Stieltjes-Wigert polynomials \cite{szeg1939orthogonal}, which are essentially a type of $q$-deformed Hermite polynomial.\par 
A Riemannian log-normal probability distribution was already introduced in \cite{schwartzman2006random} (see \cite{schwartzman2016lognormal} for a review) but is not of the random matrix type and not comparable to \eqref{ZSW} due to an extra factor in the Vandermonde part.

Another way to see this $q$-deformed structure at play is to define the $\mathsf{q}$-parameter 
\begin{equation}
\label{eq:otherq}
	\mathsf{q}= e^{- \sigma }
\end{equation}
and introduce the $\mathsf{q}$-number 
\begin{equation}
 [ x ]_{\mathsf{q}} = \frac{ \mathsf{q}^{- x/2} -  \mathsf{q}^{x/2} }{  \mathsf{q}^{- 1/2} -  \mathsf{q}^{1/2}} ,
\end{equation}
which reduces to an ordinary real number, $[ x ]_{\mathsf{q}} \to x$, in the limit $\mathsf{q} \to 1^{-}$. Then we have 
\begin{equation}
\label{ZqGUE}
z_{\beta} (\sigma) = \left\lvert 2 \sinh \frac{\sigma}{2} \right\rvert^{\frac{\beta}{2} m (m-1)} \frac{1}{m!} \int_{\R^m} \prod_{1 \le i <  \le m} \lvert [\tilde{r}_i - \tilde{r}_j]_{\mathsf{q}} \rvert^{\beta} \prod_{i=1}^{m} e^{- c_{\beta} \tilde{r}_i ^2 } \frac{ d \tilde{r}_i}{\sqrt{\pi}} 
\end{equation}
where $\tilde{r}_i = r_i / \sigma $. This latter form shows that the partition function $z_{\beta} (\sigma)$ is, up to the explicitly known overall $\sigma$-dependent factor, the $\mathsf{q}$-deformation of the Gaussian $\beta$-ensemble, for every $\beta>0$, with $\mathsf{q}$ as in \eqref{eq:otherq}.\par

We study the partition function $z_{\beta} (\sigma)$. Before that, we briefly review the relation between random matrix ensembles and symmetric spaces in Subsection \ref{sec:symspacesRMT}. Then we discuss first the case $\beta=2$ in the rest of the present Section. The other two cases $\beta=1$ and $\beta=4$ are studied in Section \ref{sec:beta-14}.\par

\subsection{Cartan classification: from symmetric spaces to matrix ensembles}
\label{sec:symspacesRMT}
A correspondence between symmetric spaces and random matrix ensembles was established by Altland and Zirnbauer \cite{AltlandZirnbauer:96} and used to classify the latter ones in terms of the former (see \cite{ZirnbauerReview,CMReview} for an overview).\footnote{These results have been recently extended to include double-coset spaces \cite{edelman2020generalized}. That setup seems especially well-suited for applications in statistical analysis.}\par
As we have seen above, when dealing with invariant joint probability distributions, the integrals only depend on the eigenvalues $(x_1, \dots, x_m)$ of the random matrix. The Jacobian coming from this reduction is the Vandermonde determinant \cite{MehtaBook}:
\begin{equation}
\label{appeq:VdM}
    J (x) = \prod_{1 \le i <j  \le m} \vert x_i - x_j \vert^{\beta} .
\end{equation}
For matrix ensembles of the transfer matrix type (following the terminology of \cite{CMReview}) the Jacobian is naturally 
\begin{equation}
\label{eq:HypVdM}
    J (x) = \prod_{1 \le i<j \le m }\, \left\lvert \sinh\left( \frac{x_{i}-x_{j}}{2} \right) \right\rvert^{\beta} 
\end{equation}
which is recast in the form \eqref{appeq:VdM} with an exponential change of variables. For circular ensembles, on the other hand, one gets \cite{Dyson3} 
\begin{equation}
\label{eq:circVdM}
    J (x) = \prod_{1 \le i<j \le m }\, \left\lvert 2 \sin \left( \frac{\theta_{i}-\theta_{j}}{2}\right) \right\rvert^{\beta} 
\end{equation}
which is related to \eqref{appeq:VdM} through $x=e^{i \theta}$. One readily observes that \eqref{appeq:VdM}, \eqref{eq:HypVdM} and \eqref{eq:circVdM} are precisely the Jacobians to pass to spherical coordinates in symmetric spaces of zero, negative and positive curvature, respectively. 
\par
We are interested in ensembles of real symmetric, Hermitian or quaternion real Hermitian matrices, corresponding to $\beta=1$, $\beta=2$ and $\beta=4$, respectively \cite{MehtaBook,ForresterBook}. These three matrix algebras are realised as the tangent space at the origin of coset spaces of Cartan type $A$, yielding the classification summarized in Table \ref{tab:CartanDyson}.\par
\begin{table}[htb]
    \centering
    \begin{tabular}{c|c|c|c|c}
         Root system & Cartan class. & matrix type & Dyson index & symmetry  \\
         \hline
         $A_{m-1}$ & A & Hermitian & $\beta=2$ & unitary \\
         $A_{m-1}$ & AI & real symmetric & $\beta=1$ & orthogonal \\
         $A_{m-1}$ & AII & quaternion Hermitian & $\beta=4$ & symplectic \\
         \hline
    \end{tabular}
    \vspace{0.3cm}
    \caption{$A$-series of the Cartan classification of random matrix ensembles.}
    \label{tab:CartanDyson}
\end{table}\par
In conclusion, we single out three hyperbolic symmetric spaces, which correspond to random transfer matrix ensembles with $\beta \in \left\{ 1,2,4 \right\}$. All of them are discussed in the following. Besides, we can go beyond Cartan type $A$: two examples of type $D$ are provided in Appendix \ref{app:SoandSp}.

\subsection{Moments of the log-normal distribution and the Stieltjes-Wigert orthogonal polynomials}
\label{sec:beta2}

In what follows we exploit the rewriting \eqref{ZSW}, which allows to directly apply standard results from random matrix theory. Define the inner product 
\begin{equation}
\label{eq:SWinnerprod}
    (f,g)_2  = \int_0 ^{\infty} \frac{dx}{\sqrt{\pi \sigma^2}} e^{- (\log x)^2 /\sigma^2 } f (x) g(x) 
\end{equation}
for analytic real functions $f$ and $g$. We keep the dependence on $\sigma$ implicit in the notation. The inner product of two monomials is\footnote{The same moments can be obtained working on the unit circle, if the function is instead Jacobi's third theta function, upon a formal replacement $q \mapsto q^{-1}$. This leads to a unitary matrix model with analogous properties to the models here discussed, see \cite{romo2012unitary,giasemidis2014torus,takahashi2014oscillatory,garcia2020matrix} for details.}

\begin{align}
    ( x^k, x^{\ell} )_2  & = \int_0 ^{\infty} \frac{dx}{\sqrt{\pi \sigma^2}} e^{- (\log x)^2  /\sigma^2 } x^{k + \ell} \notag \\
    &= \int_{- \infty} ^{+ \infty} \frac{dr}{\sqrt{\pi \sigma^2}} e^{ - r^2 /\sigma^2 + r (k+\ell+1)} =  e^{\frac{\sigma^2}{4} (k + \ell + 1)^2} . \label{mombeta2}
\end{align}

For the cases $\beta=1$ and $\beta=4$ we will need to define skew-symmetric products, but for $\beta=2$, $z_{\beta} (\sigma)$ can be evaluated exactly \cite{Forrester:Vicious}, thanks to the Stieltjes-Wigert polynomials, which are orthonormal polynomials with respect to inner product $( \cdot, \cdot)_2 $ defined in \eqref{eq:SWinnerprod}, that is 
\begin{equation}
    \left( P_k, P_{\ell} \right)_2  = \delta_{k \ell} .
\end{equation}
This family of polynomials is given by 
\begin{equation}
    P_n (x) =  (-1)^n q^{\frac{1}{2} \left(  n+ \frac{1}{2} \right) } \prod_{j=1} ^{n} \left(  1- q^{j} \right)^{- \frac{1}{2}} \sum_{\nu=0} ^{n} \left[ \begin{matrix}  n \\ \nu \end{matrix} \right]_{q} q^{\nu^2 + \frac{\nu}{2}} (-x)^{\nu} ,
\end{equation}
where the $q$-parameter is 
\begin{equation}
\label{qSW}
    q= e^{- \sigma^2 /2}
\end{equation}
(not to be confused with $\mathsf{q}$ in \eqref{eq:otherq}), and 
\begin{equation}
    \left[ \begin{matrix}  n \\ \nu \end{matrix} \right]_{q} = \frac{ [n]_q}{ [n - \nu]_q [\nu]_q } = \prod_{j=1} ^{\nu} \frac{  1- q^{n - j +1}}{1 - q^{j}}
\end{equation}
is the $q$-binomial. Let us denote by $a_n$ the leading coefficient, that is $P_n (x)= a_n x^n + \dots$, and define the \emph{monic} orthogonal polynomials $p_n (x) $, which satisfy the orthogonality relation 
\begin{equation}
    \left( p_k, p_{\ell} \right)_2  = \frac{1}{a_n ^{2}} \delta_{k \ell} ,
\end{equation}\par
with $a_0 ^{-2} = 1/q$ and 
\begin{equation}
    a_n ^{-2} = \frac{ \prod_{j=1}^{n} (1-q^j)}{ q^{2n^2 +2n +\frac{1}{2}}} 
\end{equation}
for $n =1,2, \dots$ \cite{Forrester:Vicious}.\par
The main tool to evaluate $z_{\beta=2} (\sigma)$ exactly is the celebrated Andr\'{e}ief identity \cite{Andreief,ForresterMeet}, which allows to rewrite the matrix model \eqref{ZSW} as a determinant:
\begin{equation}
    z_{2} (\sigma) = e^{\frac{\sigma^2 }{4 } m ^3 }  \det_{1 \le i, j \le m } \left[ \left( x^{i-1} , x^{j-1} \right)_2  \right] .
\end{equation}
In turn, the inner product of monomials inside the determinant can be replaced by $ \left( p_{i-1} , p_{j-1} \right)_2  $, so that the matrix (of which we are taking the determinant) becomes diagonal in this basis, and we obtain \cite{Forrester:Vicious} 
\begin{equation}
     z_{2} (\sigma) = e^{- \frac{\sigma^2 }{4} m^3  }  \prod_{n=0} ^{m-1} a_n ^{-2}  = e^{\frac{\sigma^2}{12} m (m^2-1)} \left( 1-e^{- \sigma^2 /2} \right)^{m (m-1)/2} \prod_{j=1} ^{m-1} \Gamma_q (j+1) ,
\end{equation}
with $q$ defined in \eqref{qSW} and $\Gamma_q$ the $q$-Gamma function
\begin{equation}
    \Gamma_q (j+1) = \prod_{n=1} ^{j} \left[ n \right]_{q}  .
\end{equation}
Explicitly:
\begin{equation}
     z_{2} (\sigma) = e^{\frac{\sigma^2}{12} m (m^2-1) - \frac{\sigma^2}{8} m (m-1)  }  \prod_{j=1} ^{m}  \left[ - 2  \sinh \left( \frac{\sigma^2}{4} j \right) \right]^{m-j} .
\end{equation}

\begin{figure}[thb]
    \centering
    \includegraphics[width=0.5\textwidth]{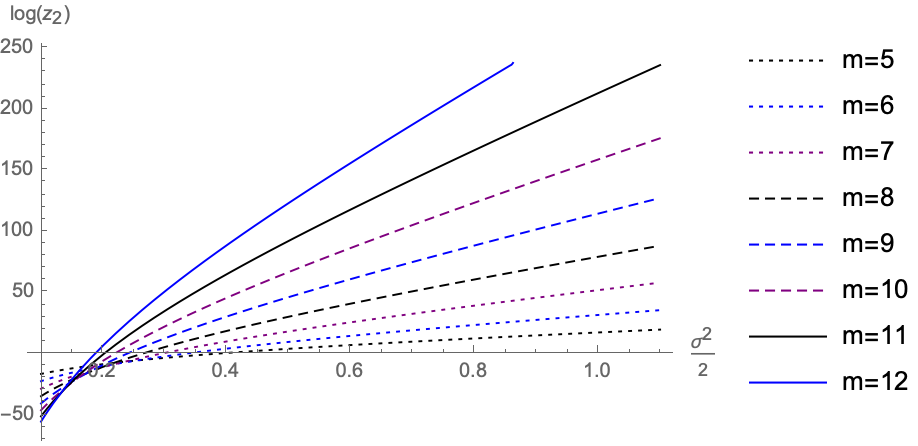}
    \caption{Plot of $\log z_{\beta=2} (\sigma)$ as a function of $\sigma^2 /2$, for $m=5,6, \dots, 12$.}
    \label{fig:z2vsSigma}
\end{figure}

\begin{figure}[thb]
    \centering
    \includegraphics[width=0.5\textwidth]{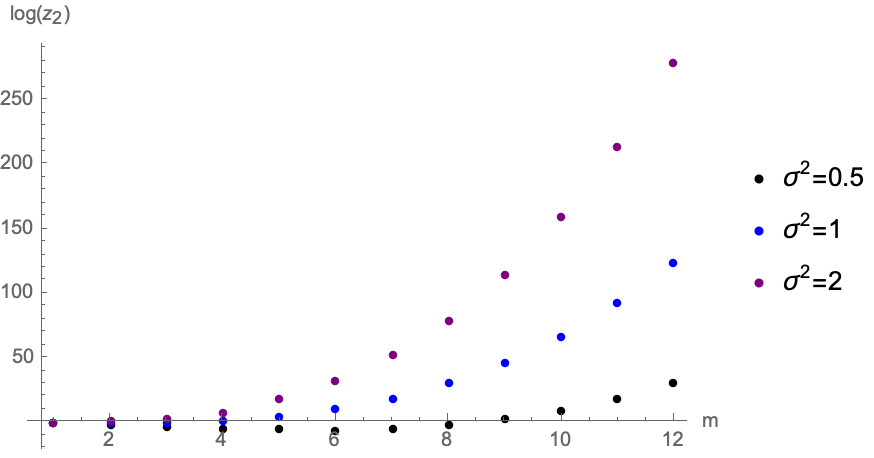}
    \caption{Plot of $\log z_{\beta=2} (\sigma)$ as a function of $m$, for fixed values of $\sigma^2=0.5,1,2$.}
    \label{fig:z2vsm}
\end{figure}

We plot $z_2 (\sigma)$ as a function of $\sigma$ for various $m$ in Figure \ref{fig:z2vsSigma}, and as a function of $m$ for various fixed $\sigma$ in Figure \ref{fig:z2vsm}.\par
\par
\par

The dependence on $m$ is better captured by $\log z_{\beta}$ rather than $z_{\beta}$ itself. For this reason, while our discussion will focus on $z_{\beta}$, in the plots and numerical evaluations we will instead consider $\log z_{\beta}$.\par

\subsection{Eigenvalue density, $\beta=2$}
\label{sec:rhobeta2}

\begin{figure}[th]
    \centering
    \includegraphics[width=0.4\textwidth]{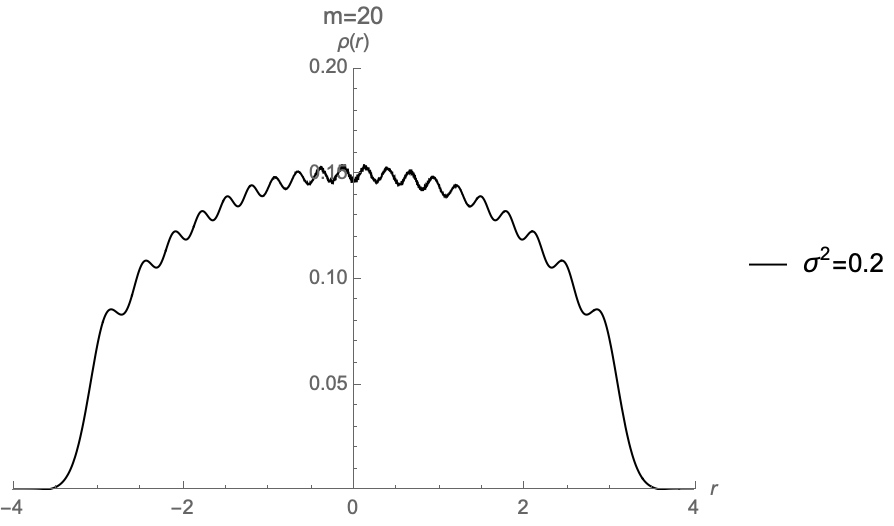}\hspace{0.04\textwidth}
    \includegraphics[width=0.4\textwidth]{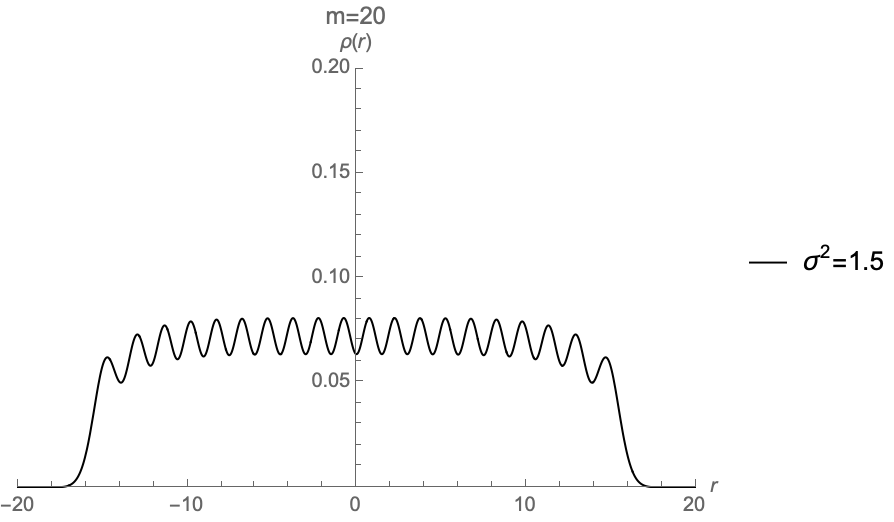}
    \caption{Eigenvalue density $\rho_2 (r; \sigma)$ at $m=20$. Left: $\sigma^2=0.2$. Right: $\sigma^2=1.5$.}
    \label{fig:rhom20}
\end{figure}

\begin{figure}[th]
    \centering
    \includegraphics[width=0.4\textwidth]{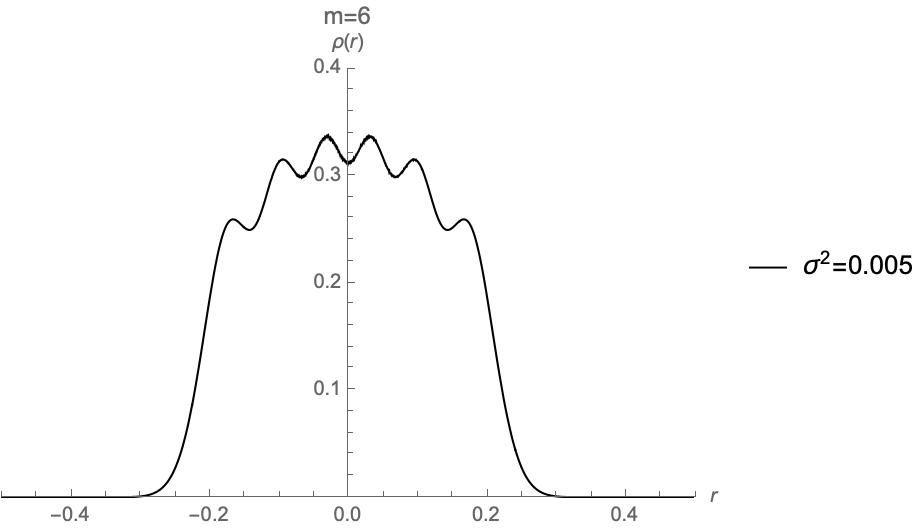}\hspace{0.04\textwidth}
    \includegraphics[width=0.4\textwidth]{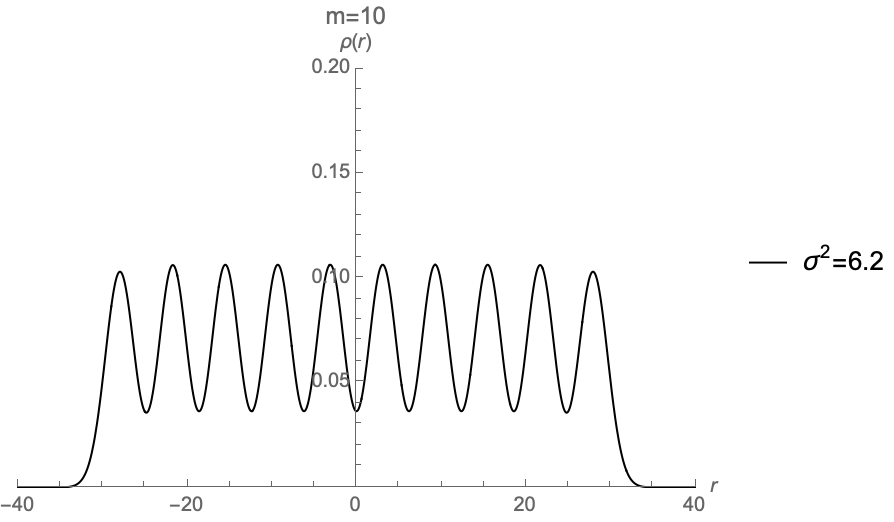}
    \caption{Eigenvalue density $\rho_2 (r ; \sigma)$. Left: small $\sigma$ regime, $m=6$ and $\sigma^2 = 0.005$. Right: large $\sigma$ regime, $m=10$ and $\sigma^2 = 6.2$. Note that the height of the vertical axis in the left picture it twice the height in the other plots.}
    \label{fig:rhosmalllarge}
\end{figure}

Let us introduce the eigenvalue density, which by definition is obtained integrating the joint probability density over $m-1$ of the $m$ eigenvalues \cite{MehtaBook}: 
\begin{equation}
\label{eq:defrho}
    \rho_{\beta} (r ; \sigma) = \frac{ m!}{(m-1)!  z_{2} (\sigma)} \cdot \frac{ e^{-  \frac{c_{\beta} }{\sigma^2} r^2 }}{ \sqrt{\pi \sigma^2} }   \cdot \int_{\R^{m-1}}\prod_{1 \le i < j \le m} \left[ 2 \sinh \left( \frac{ \vert r_i - r_j \vert }{2} \right) \right]^{\beta} \prod_{i=1} ^{m-1} e^{-  \frac{c_{\beta} }{\sigma^2} r_i^2 } \frac{d r_i}{\sqrt{\pi \sigma^2}} .
\end{equation}
Clearly, from the invariance of the integral $z_2 (\sigma)$ under permutation of the $r_i$'s, we can integrate over any $m-1$ of the $m$ variables and obtain the same definition of $\rho_{\beta} (r ; \sigma)$. Throughout this Subsection we consider $\beta=2$. It is convenient to use the change of variables $x_i = e^{r_i - \frac{\sigma^2}{2} m}$ as in \eqref{ZSW}, and exploit the Stieltjes-Wigert polynomials \cite{Tierz:02}. We can use the Christoffel-Darboux formula to rewrite the eigenvalue density of any random matrix ensemble as \cite[Eq. (5.13)]{ForresterBook}
\begin{equation}
\label{eq:CDrho}
    \rho_2 (x) = a_{m-1} ^2 w (x) \left[ p_{m-1} (x) p^{\prime} _m (x) - p_m (x) p^{\prime} _{m-1} (x) \right] ,
\end{equation}
where $w(x)$ is the weight function, $p_n$ are the monic orthogonal polynomials with respect to the weight $w(x)$, and the prime means derivative with respect to $x$. In our case, $w(x)= \frac{e^{- (\log x)^2 / \sigma^2}}{\sigma \pi^{1/2}}$ and $p_n$ are the monic Stieltjes-Wigert polynomials introduced in the previous Subsection. We find 
\begin{equation}
    p_{m-1} (x) p^{\prime} _m (x) - p_m (x) p^{\prime} _{m-1} (x) = q^{2 m^2 -m + \frac{1}{2}} \sum_{k=0}^{m-1} \sum_{\ell=0}^{m} \left\{ k (-1)^{k + \ell} x^{k + \ell -1} q^{k^2 + \ell^2 +\frac{k+ \ell }{2} } \left[ \begin{matrix} m-1 \\ k \end{matrix}\right]_q \left[ \begin{matrix} m \\ \ell  \end{matrix}\right]_q  - (k  \leftrightarrow \ell  ) \right\} .
\end{equation}
Undoing the change of variables, using $\rho_2 (x) dx = \rho_2 (e^{r + \sigma^2 m /2} ) e^{r + \sigma^2 m /2} dr$ and multiplying the last expression by the Gaussian prefactor coming from the weight function in the Christoffel-Darboux formula \eqref{eq:CDrho}, we find that $\rho_2 (r; \sigma)$ takes the form 
\begin{equation}
    \rho_2 (r; \sigma) = \sum_{k=0} ^{2m-2} c_k (q) e^{- \frac{1}{\sigma^2} \left(r - \frac{ \sigma^2}{2} (k+1-m) \right)^2  } 
\end{equation}
with the coefficients $c_k (q)$ obtainable from the expressions above. The upshot is that $\rho_2 (r ; \sigma)$ is the sum of $2m-1$ Gaussian distributions, centered at the points 
\begin{equation}
    r_k = \frac{ \sigma^2}{2} (k+1-m) , \quad k \in \left\{ -m +1, \dots, 0, \dots, m-1 \right\} .
\end{equation}
It is also easy to check that the coefficients $c_k (q)$ have alternating sign, thus the Gaussians centered at $r_k$ with $k$ odd interfere constructively with the other Gaussians at odd positions, and destructively with those at even positions. We therefore expect $\rho_2 (r ; \sigma)$ to be described by $m$ peaks and $m-1$ valleys among them. Moreover, the support of the eigenvalue density grows linearly with the product $t \equiv m \sigma^2$. We plot the eigenvalue densities for various $\sigma$ in Figure \ref{fig:rhom20}, and in the small $\sigma$ and large $\sigma$ regime in Figure \ref{fig:rhosmalllarge}.

\subsection{Limits of $z_{\beta} (\sigma)$}
\label{sec:limits}

Before delving into a more detailed analysis of $z_{\beta} (\sigma)$ for $\beta \in \left\{ 1, 4 \right\}$, it is instructive to analyze various limits for generic $\beta>0$. 

\subsubsection{$\sigma \to 0^{+}$ limit}

First, it is clear from \eqref{ZqGUE} that sending $\sigma \to 0^{+}$ we recover the usual Gaussian $\beta$-ensemble, whose partition function in known exactly for every $\beta>0$, see \cite[Eq. (1.160)]{ForresterBook}. In particular, with our normalization we have 
\begin{equation}
\label{eq:z1sigmasmall}
   z_{\beta=1} (\sigma) \xrightarrow{ \ \sigma \to 0^{+} \ } \sigma^{m (m-1)/2} 2^{m/2} \prod_{j=0} ^{m/2 -1} \frac{ \Gamma (2j+1)}{2^{2j}}
\end{equation}
for $m$ even (and a similar expression for $m$ odd), and 
\begin{equation}
    z_{\beta=2} (\sigma) \xrightarrow{ \ \sigma \to 0^{+} \ }  \frac{\sigma^{m (m-1)} }{m!} \prod_{j=1} ^{m} \frac{ \Gamma (j+1)}{2^{j-1}}
\end{equation}
\begin{equation}
    z_{\beta=4} (\sigma) \xrightarrow{ \ \sigma \to 0^{+} \ }   \frac{\sigma^{2m (m-1)} }{m! }\prod_{j=1} ^{m} \frac{ \Gamma (2j+1)}{2^{2j+1}} .
\end{equation}
Comparing with formula \eqref{eq:ztozeta}, the limit \eqref{eq:z1sigmasmall} shows in particular that $\zeta (\sigma)$ goes to zero as $\sim \sigma ^{m^2 /2}$ in the small $\sigma$ limit, that is, small variance in the dataset.\par

\subsubsection{$\sigma^2 \to \infty $ limit}
Another possible limit is the $\sigma^2 \to \infty $ limit. We give the result taking first the large $\sigma^2$ limit and then approximating for large $m$. It is convenient to write $z_{\beta} (\sigma)$ as 
\begin{equation}
\label{eq:zbetalargesigma}
    z_{\beta} (\sigma) \approx \frac{ \sigma^m}{\pi ^{m/2} m!} \int_{\R^m} \prod_{i=1} ^{m} d\tilde{r}_i ~\exp \left( - c_{\beta} \sigma^2 \sum_{i=1} ^{m} \tilde{r}_i ^2  + \sigma^2 \frac{\beta}{2} \sum_{ i \ne j}  \left\lvert \tilde{r}_i - \tilde{r}_j \right\rvert \right) ,
\end{equation}
where we have changed variables $\tilde{r}_i = r_i / \sigma^2$ and approximated $ \left[ 2 \sinh \left(  \sigma^2\frac{ \vert  x \vert }{2} \right)  \right] \approx e^{\sigma^2 \vert x \vert /2 }$ at large $\sigma^2$. Making the ansatz that $\tilde{r}_i$ grows as $\tilde{r}_i = m^{\alpha} s_i $ for some $\alpha>0$ and $s_i $ of order 1, we see that a saddle point exists in the large $m$ limit if the two terms in the exponential in \eqref{eq:zbetalargesigma} are of the same order in $m$. The first term goes as $m^{1 + 2 \alpha}$ and the second as $m^{2 +\alpha}$, meaning that the large $m$ limit requires $m^{1 + 2 \alpha} = m^{2 +\alpha}$, that is $\alpha=1$. This implies that $\log z_{\beta} (\sigma) \propto \sigma^2 m^3$ at leading order at large $\sigma^2$ and large $m$.\par 

We can also foresee the form of the sub-leading correction. Indeed, it will come from integrating over fluctuations around the saddle point. The entries of the Hessian matrix from \eqref{eq:zbetalargesigma} are $\sim c_{\beta} \sigma^2$, and performing the $m$ Gaussian integrals and taking into account the coefficient we expect this sub-leading correction to be of order $\sim m \log \sigma ^2$. Note, however, that \eqref{eq:zbetalargesigma} as it stands is not suitably written to study the saddle point equation. To get rid of the absolute value, we should restrict to the principal Weyl chamber in $\R^m$ and look for a saddle point therein. We do not pursue this approach, and instead quote the result found in \cite{GST}, where a rigorous approach to this limit has been undertaken, based on work by Baxter \cite{Baxter:63}.\footnote{The extension of Baxter's result \cite{Baxter:63} to arbitrary $\beta$, as we need, is straightforward. The statistics of extreme eigenvalues in this regime has been recently studied in \cite{Dhar:2017grt,Dhar:2018}.} One gets: 
\begin{equation}
\label{eq:zbetaBaxter}
     \log z_{\beta} (\sigma) \approx \frac{\beta}{24} \sigma^2 m (m^2-1) + \frac{\beta}{8} m \log \sigma^2 .  
\end{equation}

\subsubsection{Planar limit}

Another useful limit is the scaled large $m$ limit with $m \to \infty$ and $\sigma^2 \to 0$ keeping their product $m \sigma^2 \equiv t $ fixed. In this limit, the leading contribution to $z_{\beta} $ comes from the saddle point configuration, and one finds 
\begin{equation}
\label{eq:largeNtHooft}
    \log z_{\beta} (\sigma) \approx - \log (m!) - \frac{m}{2} \log (\pi \sigma^2 ) + \frac{\beta}{2} F_{\text{univ.}} (t) ,
\end{equation}
where $F_{\text{univ.}} (t) $ is a $\beta$-independent quantity, explicitly found solving the saddle point equation. To obtain \eqref{eq:largeNtHooft} we have used the normalization $c_{\beta}= \beta/2$, and the normalization \eqref{normcbeta} is recovered simply shifting $t \mapsto 2t$ in the $\beta=4$ case.\par
Both $F_{\text{univ.}} (t) $ and the limiting eigenvalue density $\rho (r;t) $ are explicitly known: they have been computed almost twenty years ago in the context of Chern-Simons theory \cite{Aganagic:2002wv}. We refer to \cite{Aganagic:2002wv} or the review \cite{Marino:2004eq} for detailed analysis and formulas in this regime.

In \cite{Forrester:Vicious} this scaled limit was studied and subdivided in three cases, according to the value of $m \sigma^2 \equiv t $ being finite, infinite or $0$. In Section \ref{sec:diffusion}, we will discuss the relationship between the probability densities in \cite{cheng2013novel,said2017riemannian,said2017gaussian} with diffusion processes and how they appear in the study of Brownian motion on Weyl chambers of Lie groups.


\section{The cases of real symmetric and quaternion Hermitian matrices}
\label{sec:beta-14}

According to the results in Subsection \ref{sec:limits}, we can extract the behaviour of $z_{\beta} (\sigma) $ in certain limits from the knowledge of $z_{2} (\sigma) $. Nevertheless, we can do more, and evaluate exactly $z_{1} (\sigma)$ and $z_{4} (\sigma)$ using standard tools in random matrix theory.

\subsection{Skew-symmetric products and $\beta=1$ partition function}
\label{sec:beta1}

We discuss $z_{\beta=1} (\sigma)$, which is directly related to $\zeta (\sigma)$ through \eqref{eq:ztozeta}. 
In contrast to the case $\beta=2$, the partition functions $z_{\beta=1} (\sigma)$ and $z_{\beta=4} (\sigma)$ do not correspond to determinants, but rather to Pfaffians. It is convenient to discuss separately the two cases with $m$ even or $m$ odd, starting with the former. Before that, we define the skew-symmetric products \cite{AFNvM} 
\begin{equation}
    \langle f, g \rangle_{4}  = \frac{1}{2} \int_0 ^{\infty} \frac{dx}{\sqrt{\pi \sigma^2}} e^{- (\log x)^2 /\sigma^2 } \left[  f (x) g^{\prime } (x) - f^{\prime } (x) g(x) \right] ,
\end{equation}
where $f^{\prime} = \frac{df}{dx}$, and 
\begin{equation}
    \langle f, g \rangle_{1}  = \frac{1}{2} \int_0 ^{\infty} \frac{dx}{\sqrt{\pi \sigma^2}} e^{- (\log x)^2 /2 \sigma^2 } f (x) \int_0 ^{\infty} \frac{dy}{\sqrt{\pi \sigma^2}} e^{- (\log y)^2 /2 \sigma^2 } g (y)  \text{sign} (x-y) .
\label{eq:skewprod1SW}
\end{equation}
Note that the coefficient in the exponent of the weight function is $c_{\beta}$ as defined in \eqref{normcbeta}.\par
Computing the skew-symmetric products of any pair of monomials gives 
\begin{align}
     \langle x^{k}, x^{\ell} \rangle_{4}  &= \frac{1}{2} \int_{- \infty} ^{+ \infty} \frac{dr}{\sqrt{ \pi \sigma^2}} e^{- r^2 /\sigma^2 +r} \left[ \ell e^{k + \ell -1} - k e^{k+ \ell -1} \right] \notag \\
     &= e^{\frac{ \sigma^2}{4} (k + \ell)^2} ~ \frac{ ( \ell - k )}{2}  \label{skewmombeta4}
\end{align}
and 
\begin{align}
 \langle x^{k}, x^{\ell} \rangle_{1}  &= \frac{1}{2} \int_{- \infty} ^{+ \infty} \frac{dr_1}{\sqrt{ \pi \sigma^2}} e^{- r_1 ^2 /2 \sigma^2 +r_1 (k +1) } \left[ \int_{- \infty} ^{r_1} \frac{dr_2}{\sqrt{ \pi \sigma^2}} e^{- r_2 ^2 /2 \sigma^2 +r_2 (\ell +1) }  -   \int_{r_1} ^{+ \infty} \frac{dr_2}{\sqrt{ \pi \sigma^2}} e^{- r_2 ^2 /2 \sigma^2 +r_2 (\ell +1) }    \right]  \notag \\
 & = e^{\frac{\sigma^2}{2} (\ell +1)^2 }  \int_{- \infty} ^{+ \infty}  \frac{dr_1}{\sqrt{ 2 \pi \sigma^2}} e^{- r_1 ^2 /2 \sigma^2 +r_1 (k +1) } \mathrm{erf} \left(  \frac{r_1}{\sqrt{2 \sigma^2}} + \sqrt{\frac{\sigma^2}{2}} (\ell +1)  \right) \notag \\
 &= e^{\frac{\sigma^2}{2} \left[ (k+1)^2 + (\ell +1)^2 \right] } ~ \mathrm{erf} \left(\frac{\sigma}{2} ( \ell - k )  \right) , \label{skewmombeta1}
\end{align}
where we have used the change of variables $x=e^{r}$. In \eqref{skewmombeta1}, $\mathrm{erf}(x)$ is the error function, and in the last line we have used the integration formula for $\mathrm{erf}(x)$ with a Gaussian weight. The error function is odd, $\mathrm{erf}(-x)=- \mathrm{erf}(x)$, whence both products are manifestly skew-symmetric.

Assuming $m$ even, we use the de Bruijn identity \cite{deBruijn} to write 
\begin{equation}
    z_{\beta=1} (\sigma) \vert_{m \text{ even}} = e^{- \frac{ \sigma^2}{8} m (m+1)^2} \underset{1 \le i,j \le m}{\mathrm{Pf}} \left[ 2 \langle x^{i-1}, x^{j-1} \rangle_1   \right] 
\label{eq:Zbeta1Pfaff}
\end{equation}
with the skew-symmetric product $\langle x^{i-1}, x^{j-1} \rangle_1 $ given in \eqref{skewmombeta1}. For $m$ odd, instead, the de Bruijn identity \cite{deBruijn} leads us to the Pfaffian of a $(m+1) \times (m+1)$ skew-symmetric matrix: 
\begin{equation}
  z_{\beta=1} (\sigma) \vert_{m \text{ odd}} = e^{ -\frac{ \sigma^2}{8} m (m+1)^2} \underset{1 \le i,j \le m+1}{\mathrm{Pf}} \left[ \begin{matrix}  \left[  2 \langle x^{i-1}, x^{j-1} \rangle_1  \right]_{i,j \le m} & \left[ 2 \left( 1, x^{i-1} \right)_2  \right]_{i \le m} \\ \left[ - 2 \left( 1, x^{j-1} \right)_2  \right]_{j \le m}  & 0 \end{matrix}  \right] .
 \end{equation}
 In both cases, we could expand the Pfaffian and obtain $z_{\beta=1} (\sigma)$, and hence $\zeta (\sigma)$, as a finite sum of terms, explicitly known from \eqref{skewmombeta1}. The advantage of this approach is that the Pfaffians can be obtained explicitly for fixed $m$ with the aid of a computer algebra. We provide numerical results at various $m$ and $\sigma^2$ in Tables \ref{tab:zbeta1} and \ref{tab:zmevenbeta1}, computed in \textsc{Mathematica} using the algorithm of \cite{Wimmer}. \par
 
 \begin{small}
\begin{table}[ht]
\centering
\hspace*{-1cm}
\begin{tabular}{c ||c|c|c|c|c|c|c|c|c|c|c|c| }
$ m ~ \vert ~ \sigma^2  $ & 0.1 & 0.2 & 0.3 & 0.4 & 0.5 & 0.6 & 0.7 & 0.8 & 0.9 & 1.0 & 1.1 & 1.2  \\
\hline 
2&  -0.680 & 0.376 & 1.001 & 1.449 & 1.801 & 2.092 & 2.340 & 2.557 & 2.751 & 2.926 & 3.086 & 3.234 \\
3 &  -2.776 & -0.033 & 1.561 & 2.685 & 3.553 & 4.260 & 4.855 & 5.370 & 5.824 & 6.230 & 6.597 & 6.933 \\
4 &  -2.173 & 1.446 & 3.628 & 5.223 & 6.495 & 7.566 & 8.497 & 9.328 & 10.080 & 10.773 & 11.416 & 12.020  \\
5 &  -6.142 & 0.537 & 4.492 & 7.339 & 9.583 & 11.449 & 13.057 & 14.478 & 15.758 & 16.930 & 18.014 & 19.028  \\ 
6 &  -4.374 & 3.418 & 8.200 & 11.755 & 14.641 & 17.108 & 19.287 & 21.259 & 23.073 & 24.765 & 26.359 & 27.874 \\ 
7 &  -9.899 & 2.342 & 9.725 & 15.138 & 19.485 & 23.169 & 26.403 & 29.315 & 31.987 & 34.473 & 36.813 & 39.035  \\
8 &  -7.182 & 6.511 & 15.056 & 21.513 & 26.837 & 31.456 & 35.596 & 39.390 & 42.927 & 46.264 & 49.444 & 52.496  \\
9 &  -14.007 & 5.538 & 17.541 & 26.502 & 33.829 & 40.142 & 45.779 & 50.935 & 55.734 & 60.261 & 64.575 & 68.720 \\
10&  -10.489 & 10.945 & 24.546 & 34.983 & 43.719 & 51.401 & 58.374 & 64.842 & 70.936 & 76.742 & 82.323 & 87.725  \\
11 &  -18.406 & 10.309 & 28.260 & 41.895 & 53.230 & 63.156 & 72.146 & 80.477 & 88.324 & 95.805 & 103.003 & 109.975  \\
12 & -14.188 & 16.951 & 37.033 & 52.678 & 65.954 & 77.778 & 88.632 & 98.803 & 108.472 & 117.760 & 126.751 & 135.506  \\
 \hline \hline \\
 $ m ~ \vert ~ \sigma^2 $ & 1.3 & 1.4 & 1.5 & 1.6 & 1.7 & 1.8 & 1.9 & 2.0 & 2.1 & 2.2 & 2.3 & 2.4  \\
\hline 
2&   3.372 & 3.500 &  3.621 & 3.735 & 3.844 & 3.947 & 4.046 & 4.141 & 4.232 & 4.319 & 4.404 & 4.486 \\
3 &   7.243 & 7.530 & 7.799 & 8.051 & 8.290 & 8.516 & 8.731 & 8.936 & 9.133 & 9.322 & 9.504 & 9.680 \\
4 &  12.590 & 13.131 & 13.649  & 14.145 & 14.624 & 15.086 & 15.534 & 15.969 & 16.393 & 16.807 & 17.211 & 17.607 \\
5 & 19.984 & 20.891 & 21.757   & 22.588 & 23.388 & 24.162 & 24.914 & 25.645 & 26.358 & 27.055 & 27.738 & 28.409 \\ 
6 &  29.323 & 30.716 & 32.062 &  33.368 & 34.638 & 35.879 & 37.093 & 38.283 & 39.452 & 40.602 & 41.735 & 42.854 \\ 
7 &   41.161 & 43.207 & 45.186 &   47.108 & 48.981 & 50.813 & 52.608 & 54.371 & 56.107 & 57.817 & 59.505 & 61.173 \\
8 & 55.443 & 58.303 & 61.090 &  63.815 & 66.486 & 69.110 & 71.694 & 74.242 & 76.759 & 79.248 & 81.713 & 84.154 \\
9 &   72.726 & 76.619 & 80.417   & 84.135 & 87.785 & 91.377 & 94.918 & 98.4151 & 101.874 & 105.298 & 108.692 & 112.059 \\
10&  92.979 & 98.112 & 103.144 &
   108.090 & 112.964 & 117.774 & 122.529 & 127.237 & 131.902 & 136.530 & 141.126 & 145.691 \\
11 &  116.766 & 123.408 & 129.927
   & 136.343 & 142.671 & 148.924 & 155.111 & 161.243 & 167.325 & 173.364 & 179.364 & 185.329 \\
12 &  144.071 & 152.479 & 160.758
   & 168.927 & 177.002 & 184.999 & 192.926 & 200.793 & 208.607 & 216.374 & 224.101 & 231.790 \\
 \hline 
\end{tabular}
\vspace{0.3cm}
\caption{$\log \zeta (\sigma)$ for $m$ from $2$ to $12$ and $\sigma^2$ from $=0.1$ to $2.4$. Evaluation time $0.207$s (whole table) on macbook.}
\label{tab:zbeta1}
\end{table}\par

\begin{table}[ht]
\centering
\hspace*{-1cm}
\begin{tabular}{c||c|c|c|c|c|c|c|c|c|c|}
$m $ & 2 & 4 & 6 & 8 & 10 & 12 &14& 16 & 18 & 20  \\
\hline
$- \log  z_{1}$  &  4.150 & 13.822 & 29.007 & 49.694 & 71.487 & 93.286 & 113.424 & 135.129 & 156.057 & 175.890 \\
\hline \hline \\
$m $ &  22 & 24 & 26 & 28 & 30 & 32 & 34 & 36 & 38 & 40 \\
\hline
$- \log z_{1}$  & 199.322 & 219.185 & 244.263 & 267.156 & 295.836 & 319.471 & 349.711 & 378.587 & 410.486 & 444.106  \\
\hline 
\end{tabular}
\vspace{0.3cm}
\caption{$-\log z_{\beta=1}$ for even $m$ from $2$ to $40$ and fixed $\sigma^2=0.01$. Evaluation time $0.133$s (whole table) on macbook.}
\label{tab:zmevenbeta1}
\end{table}
\end{small}
\par

\begin{figure}[th]
    \centering
    \includegraphics[width=0.5\textwidth]{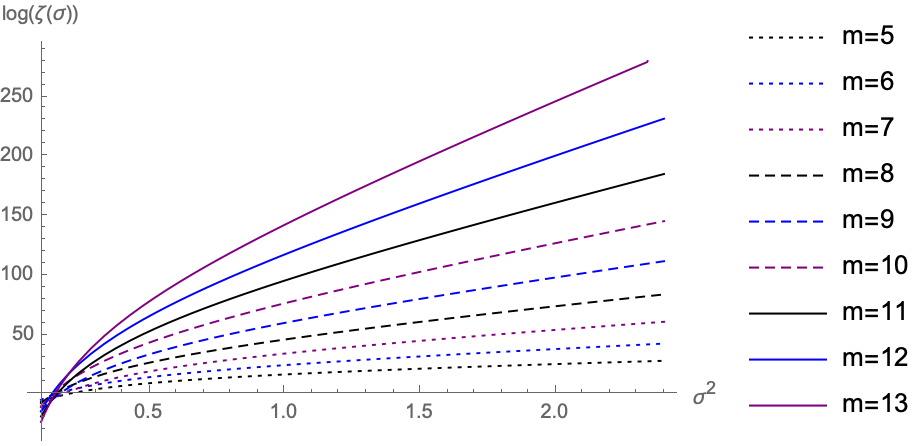}
    \caption{$\log \zeta (\sigma)$ as a function of $\sigma^2$ for various fixed values of $m$.}
    \label{fig:zetavssigma}
\end{figure}

\begin{figure}[th]
    \centering
    \includegraphics[width=0.5\textwidth]{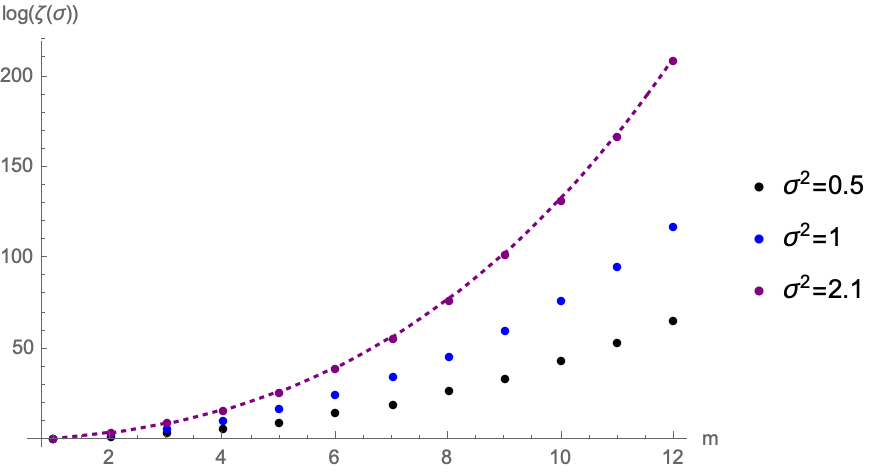}
    \caption{$\log \zeta (\sigma)$ as a function of $m$ for $\sigma^2=0.5,1,2.1$. The dashed line shows the theoretical prediction in the large $\sigma$ regime at $\sigma=2.1$.}
    \label{fig:zetavsrank}
\end{figure}

\begin{figure}[th]
    \centering
    \includegraphics[width=0.47\textwidth]{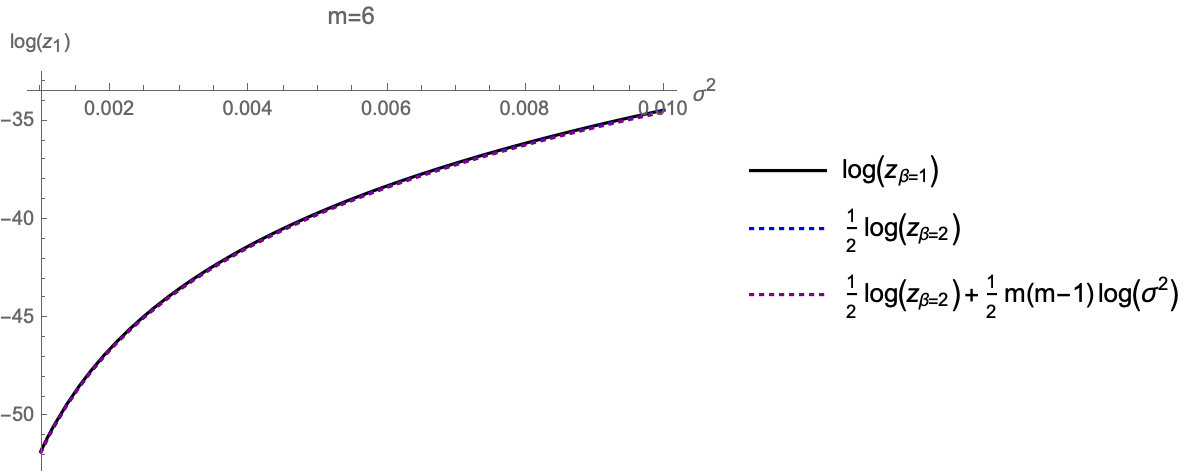}
    \hspace{0.01\textwidth}
    \includegraphics[width=0.47\textwidth]{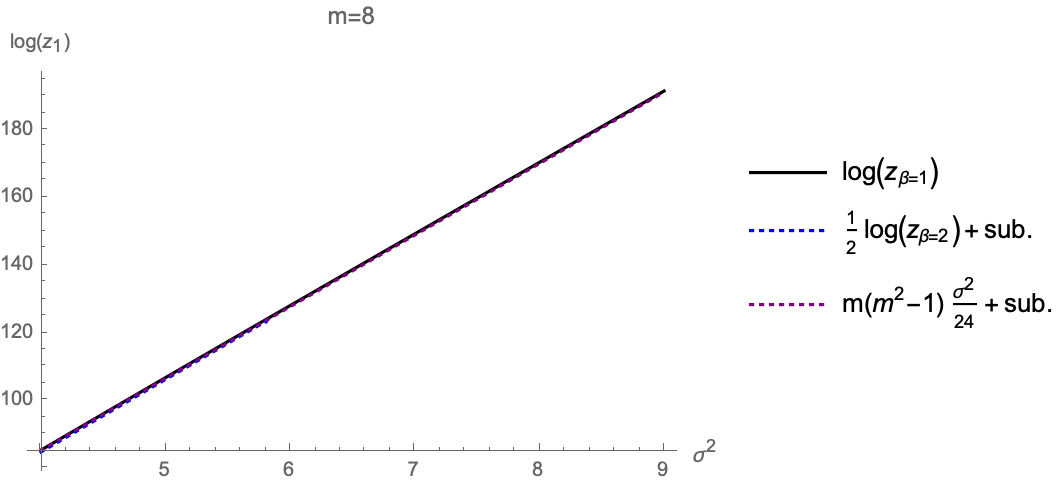}
    \caption{Left: $\log z_{\beta=1} $ in the small $\sigma$ regime, $m=6$. Right: $\log z_{\beta=1} $ in the large $\sigma$ regime, $m=8$. ``sub.'' stands for the sub-leading corrections obtained in \eqref{eq:zbetaBaxter}.}
    \label{fig:zstrongweak}
\end{figure}

From $z_{\beta=1} (\sigma)$ we immediately obtain $\zeta (\sigma)$, which we show in Figure \ref{fig:zetavssigma} as a function of $\sigma^2$ for various $m$, and in Figure \ref{fig:zetavsrank} as a function of $m$ for various fixed values of $\sigma^2$. We also show the agreement of $\log z_{\beta=1}$ with the theoretical predictions of Subsection \ref{sec:limits} for small $\sigma$ and large $\sigma$ in Figure \ref{fig:zstrongweak}.

\par
\par

\par
\medskip
While the Pfaffian representation of $z_{\beta=1} $ is very useful for computational purposes, the simple example $m=2$ can be easily solved by direct integration:
\begin{align}
    z_{\beta=1} (\sigma) \vert_{m=2} & = \frac{1}{2!} \int_{- \infty} ^{\infty} \frac{ dr_1}{\sqrt{\pi \sigma^2}} e^{- \frac{r_1 ^2 }{2 \sigma^2}} ~ \int_{- \infty} ^{\infty} \frac{ dr_2}{\sqrt{\pi \sigma^2}} e^{- \frac{r_2 ^2 }{2 \sigma^2}} ~  2 \sinh \left( \frac{ \vert r_1 - r_2 \vert}{2} \right) \notag \\
    & = \frac{1}{2} \int_{- \infty} ^{\infty} \frac{ dr_1}{\sqrt{\pi \sigma^2}} e^{- \frac{r_1 ^2 }{2 \sigma^2}} ~  \int_{- \infty} ^{\infty} \frac{dr_2}{\sqrt{ \pi \sigma^2}} e^{ - \frac{ r_2 ^2}{2 \sigma^2} } \mathrm{sgn} (r_1 - r_2) ~  \left[   e^{\frac{r_1}{2} - \frac{ r_2}{2} } -  e^{- \frac{r_1}{2}  + \frac{ r_2}{2} } \right] \notag \\
    & = e^{\frac{\sigma^2}{8}} \int_{- \infty} ^{\infty} \frac{ dr_1}{\sqrt{ 2 \pi \sigma^2}} e^{- \frac{r_1 ^2 }{2 \sigma^2}} \left[  e^{ \frac{r_1}{2}} \mathrm{erf} \left( \frac{ r_1 + \sigma^2 /2 }{ \sqrt{2 \sigma^2}}  \right) - e^{ - \frac{r_1}{2}} \mathrm{erf} \left( \frac{ r_1 - \sigma^2 /2 }{ \sqrt{2 \sigma^2}}  \right) \right]  .
\end{align}
We now observe that the integral of an error function with a Gaussian weight is again an error function and, after simplifications, we obtain 
\begin{equation}
    z_{\beta=1} (\sigma) \vert_{m=2} = 2 e^{\frac{\sigma^2}{4}} \mathrm{erf} \left( \frac{ \vert \sigma \vert }{2}  \right) .
\end{equation}
This is a check of the result \eqref{eq:Zbeta1Pfaff} for the smallest value of $m$. For higher values of $m$ a direct computation becomes more involved. However, from the Pfaffian representation, we infer that the result is always a finite sum of products of error functions, weighted by integer powers of $e^{\sigma^2 /4}$. So, for example, de Bruijn's identity immediately gives 
\begin{align}
    z_{\beta=1} (\sigma) \vert_{m=3} & = 4 \left[ e^{- \frac{5}{4} \sigma^2} \left(  1+ e^{2 \sigma^2} \right) \mathrm{erf} \left( \frac{ \vert \sigma \vert }{2} \right) -  \mathrm{erf} \left(  \vert \sigma \vert \right) \right] , \\
     z_{\beta=1} (\sigma) \vert_{m=4} & = 4 e^{\frac{5}{2} \sigma^2} \left[ \mathrm{erf} \left( \frac{ \vert \sigma \vert }{2} \right)^2 - \mathrm{erf} \left( \vert \sigma \vert \right)^2 +  \mathrm{erf} \left( \frac{ \vert \sigma \vert }{2} \right) \mathrm{erf} \left( \frac{ 3 \vert \sigma \vert }{2} \right) \right] .
\end{align}
A similar type of expression, involving a sum of error functions, is obtained for the Gaussian distribution on the Poincar\'{e} ball \cite{mathieu2019continuous}. In that case, the Vandermonde term $\sinh \left( \vert r_i - r_j \vert /2 \right)$ is replaced by the simpler term $\sinh \left( \vert r \vert /2 \right)$.

It would be interesting if the quick numerical evaluation of the normalization constant we obtained, based on Pfaffians, can be of further exploited, given the relevance of the distribution to the engineering of algorithms for machine learning and detection of structured collections of data on graphs \cite{gerald2020node}, as well as in the study of auto-encoders \cite{mathieu2019continuous,ovinnikov2020poincare}, just to name a few applications, in addition to the ones discussed in \cite{cheng2013novel,said2017gaussian,said2017riemannian,said2019warped}.

\subsection{Eigenvalue density, $\beta=1$}
\label{sec:rhobeta1}

The eigenvalue density at $\beta=2$ has been obtained relying on the explicit knowledge of the Stieltjes-Wigert polynomials. For a closed expression of $\rho_{\beta=1}$, we would need the corresponding skew-orthogonal polynomials \cite{MehtaBook,AFNvM}, which however are not known. To gain insight, let us first study the simplest case $m=2$. Then the eigenvalue density is obtained by direct integration, and the computations are identical to the ones at the end previous Subsection. We find: 
\begin{align}
    \rho_{\beta=1} (r; \sigma) \vert_{m=2} & = \frac{1}{z_{\beta=1} (\sigma) \vert_{m=2} } \int_{- \infty} ^{\infty} \frac{dr_2}{\pi \sigma^2} e^{ - \frac{(r ^2 + r_2 ^2)}{2 \sigma^2} } ~ 2 \sinh \left( \frac{\vert r - r_2 \vert}{2} \right) \notag \\
    & =  \frac{  e^{ - \frac{ \sigma^2}{8} } }{ \mathrm{erf} \left( \vert \sigma \vert /2 \right) } \cdot \frac{ e^{- \frac{r^2}{2 \sigma^2}} }{ \sqrt{ 2 \pi \sigma^2}} ~ \left[  e^{\frac{r}{2} } \mathrm{erf} \left( \frac{r + \sigma^2/2}{\sqrt{2 \sigma^2}} \right) -  e^{- \frac{r}{2} } \mathrm{erf} \left( \frac{r - \sigma^2/2}{\sqrt{2 \sigma^2}} \right)  \right] .
\end{align}
This formula already shows the salient features of the $\beta=1$ eigenvalue density: it is a sum of products of Gaussians (centered at different, shifted points) multiplied by error functions.\par

Although without the skew-orthogonal polynomials in closed form, the eigenvalue density can still be calculated for any fixed $m$. For this, we need to introduce some notation. Let $\widetilde{p}_n (x)$ be the monic skew-orthogonal polynomials of the Stieltjes-Wigert family, that is, they satisfy 
\begin{equation}
\label{skeworthprop}
    \left\langle  \widetilde{p}_{2k} , \widetilde{p}_{2 \ell+1} \right\rangle_1 = \widetilde{h}_{k} \delta_{k \ell} , \quad  \left\langle  \widetilde{p}_{2k} , \widetilde{p}_{2 \ell} \right\rangle _1= 0 =  \left\langle  \widetilde{p}_{2k+1} , \widetilde{p}_{2 \ell+1} \right\rangle_1 
\end{equation}
with the skew-symmetric product $\langle \cdot , \cdot \rangle_1$ defined in \eqref{eq:skewprod1SW}. These polynomials have an explicit Pfaffian representation \cite{AHvM:Pfaff,Chang:Pfaff}. To reduce clutter, let us denote throughout this Subsection 
\begin{equation}
    \mathbf{z}_{1} ^{(m)} = e^{\frac{\sigma^2}{8} m (m+1)^2 } ~ z_{\beta=1} ( \sigma ) \vert_{m} .
\end{equation}
That is, $\mathbf{z}_{1} ^{(m)}$ is just the partition function at $\beta=1$ for a given value of $m$, with the overall factor coming from the definition \eqref{ZSW} stripped off to avoid cumbersome coefficients.\par
The constants $\widetilde{h}_k$ in \eqref{skeworthprop} can then be expressed as $\widetilde{h}_k = \frac{ \mathbf{z}_1 ^{(2k+2)} }{ \mathbf{z}_1 ^{(2k)} }$. Moreover we have 
\begin{equation}
    \widetilde{p}_{2 \ell} (x) = \frac{1}{\mathbf{z}_{1} ^{(2 \ell)} } \underset{1 \le i,j \le 2\ell +2 }{\mathrm{Pf}} \left[ \begin{matrix}  \left[  2 \langle x^{i-1}, x^{j-1} \rangle_1  \right]_{i,j \le 2 \ell +1 } & \left( x^{i-1} \right)_{i \le 2 \ell +1}  \\ \left( - x^{j-1} \right)_{j \le 2 \ell +1}  & 0 \end{matrix}  \right] 
\end{equation}
for even degree $n= 2 \ell $, and 
\begin{equation}
    \widetilde{p}_{2 \ell +1 } (x) = \frac{1}{\mathbf{z}_{1} ^{(2 \ell+1)}  } \underset{1 \le i,j \le 2n+4 }{\mathrm{Pf}} \left[ \mathcal{M} (x)  \right] ,
\end{equation}
for odd degree $n= 2 \ell +1 $. In the latter expression we have introduced the skew-symmetric matrix \cite{Chang:Pfaff} 
\begin{equation}
    \mathcal{M} (x) = \left( \begin{matrix}  0 & 2 (1,1)_2 & 2 (1,x)_2 & 2 (1,x^2)_2 & \cdots & 2 (1,x^{2 \ell+1})_2 & 0 \\
    - 2(1,1)_2 & 0 & 2 \langle 1,x \rangle_1 & 2 \langle 1,x^2 \rangle_1 & \cdots & 1 \\
    - 2(1,x)_2 & - 2 \langle 1,x \rangle_1 & 0 & 2 \langle x, x^2 \rangle_1 & \cdots & x \\
    - 2(1,x^2)_2 & - 2 \langle 1,x^2 \rangle_1 & - 2 \langle x, x^2 \rangle_1 & 0 & \cdots & x^2 \\
    \vdots & \ddots & \ddots & \ddots & \cdots & \vdots \\
    - 2 (1, x^{2 \ell +1})_2 & - 2 \langle 1, x^{2 \ell +1} \rangle_1 & - 2 \langle x, x^{2 \ell +1} \rangle_1 & - 2 \langle x^2 , x^{2 \ell +1} \rangle_1 & \cdots & x^{2 \ell +1} \\
    0 & -1 & -x & -x^2 & \cdots & 0 
    \end{matrix}   \right) .
\end{equation}
We also need the functions 
\begin{align}
    \widetilde{\phi} _n (x) & = \frac{1}{2} \int_0 ^{\infty} \frac{dy}{\sqrt{ \pi \sigma^2}} e^{- \left( \log y  \right)^2 / 2 \sigma^2 } \widetilde{p}_n (y) ~ \mathrm{sgn} (x-y) .
\end{align}

Explicit expressions of the first few skew orthogonal polynomials and of the corresponding $\widetilde{\phi} _n (x)$ functions are given in Appendix \ref{app:skewpol}.

Putting these definitions at work, we can obtain $\rho_{\beta=1} (x)$ as a particular case of \cite[Proposition 6.3.3, Page 254]{ForresterBook}. We focus on the case of even $m$ for simplicity, but a similar expression exists for odd $m$. We get 
\begin{equation}
\label{eq:rhobeta1SWformal}
    \rho_{1} (x) = \frac{e^{- \frac{\left( \log x \right)^2 }{2 \sigma^2} } }{ \sqrt{ \pi \sigma^2 } } \sum_{k=0} ^{m/2 -1}  \frac{1}{\widetilde{h}_k } \left[  \widetilde{\phi} _{2 k } (x) \widetilde{p}_{2k+1} (x) -   \widetilde{\phi} _{2 k +1} (x) \widetilde{p}_{2k} (x) \right] .
\end{equation}
This expression allows to characterize the eigenvalue density $\rho_{\beta=1} (r; \sigma)$, undoing the change of variables $x= e^{r - \frac{\sigma^2}{2} (m+1)}$. Each $\widetilde{p}_n (x)$ is a polynomial of degree $n$ in the variable $x$ (and hence in $e^r$), with coefficients ratios of sums in which each term is of the form $e^{ \sigma^2  a /4} \mathrm{erf} (\vert \sigma \vert b /2 )$ with $a$ and $b$ integers. In turn, if we write schematically 
\begin{equation}
    \widetilde{p}_n (x) = \sum_{k=1} 
^{n} \alpha_{n;k} x^k
\end{equation}
with $\alpha_{n;n} =1$ by definition, we find 
\begin{equation}
    \widetilde{\phi}_n (e^{u} ) = \frac{1}{\sqrt{2}} \sum_{k=0} ^{n} \alpha_{n;k} e^{ \sigma^2 \frac{ (k+1)^2}{2}} ~ \mathrm{erf} \left( \frac{ u - k \sigma^2}{\sqrt{2 \sigma^2}} \right) .
\end{equation}
Writing \eqref{eq:rhobeta1SWformal} in terms of the exponential variable $r$ instead of $x$, and putting all together after some rewriting, we arrive at 
\begin{align}
    \rho_{\beta=1} (r; \sigma) & = \frac{e^{- \frac{\sigma^2}{8} (m+5)^2}  }{\sqrt{2 \pi \sigma^2}} \cdot e^{- \frac{r^2}{2 \sigma^2}  + \frac{r}{2} (m+3)}  \cdot \sum_{\ell=0} ^{m/2 -1} \frac{ \mathbf{z}_1 ^{(2\ell)}}{ \mathbf{z}_1 ^{(2\ell+2)} } \notag  \\
    & \times \sum_{j=0} ^{2 \ell} \sum_{k=0} ^{2 \ell+1}  \left[  \alpha_{2 \ell; j} \alpha_{2 \ell+1; k} e^{ k r + \frac{\sigma^2}{2} ((j+1)^2 - (m+1)k )  } ~ \mathrm{erf} \left( \frac{r - (j+m/2)\sigma^2 }{\sqrt{2 \sigma^2}} \right) \right. \notag \\
    & \qquad \left. -  \alpha_{2 \ell; k} \alpha_{2 \ell+1; j} e^{ j r + \frac{\sigma^2}{2} ((k+1)^2 - (m+1)j )  } ~ \mathrm{erf} \left( \frac{r - (k+m/2)\sigma^2 }{\sqrt{2 \sigma^2}} \right) \right] .
\end{align}
In the latter expression the coefficients $\alpha_{n;k}$ are read off form the skew-orthogonal polynomials $\widetilde{p}_n (x)$. Since the formulas for both the polynomials and the normalizations $\mathbf{z}_{1} ^{(n)}$ are known, the challenge to obtain the density $\rho_{\beta=1} (r; \sigma)$ is reduced to the task of evaluating Pfaffians. Moreover, once $\left\{ \widetilde{p}_n (e^{u}) , \widetilde{\phi}_n (e^{u}) \right\}_{n=0, \dots, m-1}$ are found, it is possible to go beyond the eigenvalue density and compute any marginal distribution of the eigenvalues \cite[Chapter 6]{MehtaBook}, \cite[Chapter 6]{ForresterBook}.\par
We conclude that  $\rho_{\beta=1} (r; \sigma)$ is a sum of terms that depend on $r$ through the product of a Gaussian and an error function in $r$. Comparing the results at $\beta=2$ and $\beta=1$, we see that each Gaussian term in $\rho_{\beta=2} (r; \sigma)$ is corrected by an error function in the $\beta=1$ case. The same happens with the coefficients: the dependence $\sim e^{\sigma^2 a /2}$ in the $\beta=2$ case is replaced by $\sim e^{\sigma^2 a /2} \mathrm{erf} (\vert \sigma \vert b /2)$ at $\beta=1$.

\subsection{$\beta=4$ partition function}
\label{sec:beta4}

\begin{figure}[th]
    \centering
    \includegraphics[width=0.5\textwidth]{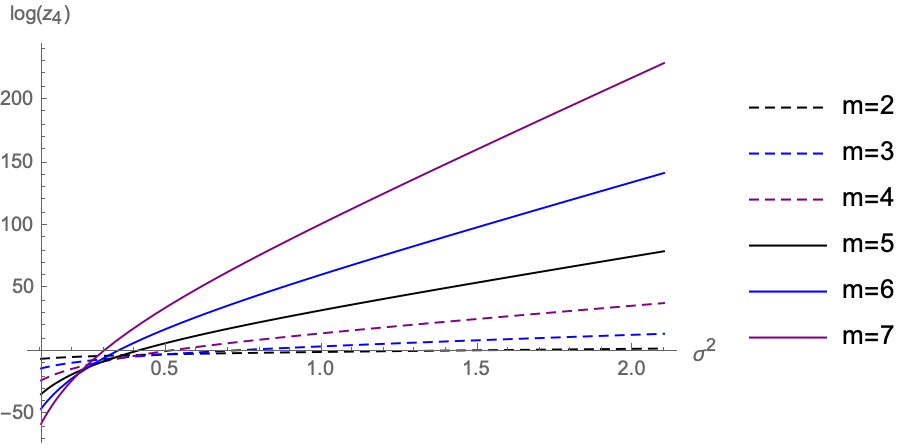}
    \caption{$\log z_{\beta=4} (\sigma)$ as a function of $\sigma^2$ for various fixed values of $m$.}
    \label{fig:z4vssigma}
\end{figure}

\begin{figure}[th]
    \centering
    \includegraphics[width=0.5\textwidth]{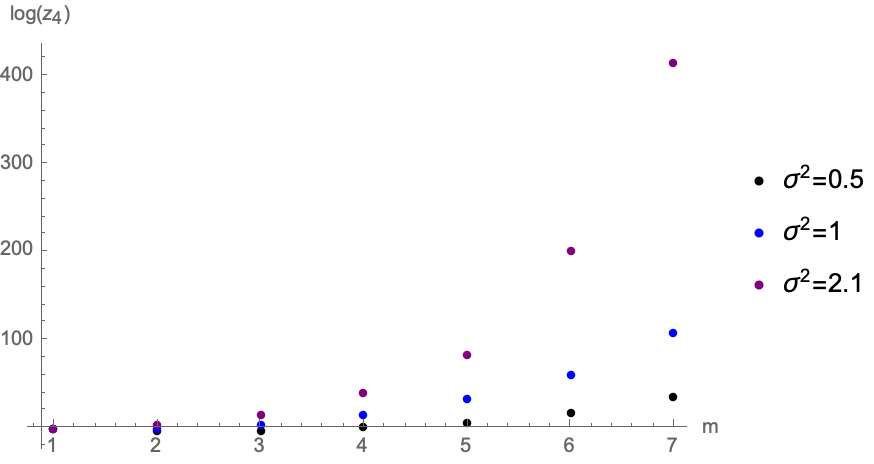}
    \caption{$\log z_{\beta=4} (\sigma)$ as a function of $m$ for $\sigma^2=0.5,1,2.1$.}
    \label{fig:z4vsrank}
\end{figure}

We can apply the tools of Subsection \ref{sec:beta1} to study the partition function $z_{\beta=4} (\sigma)$ as well. This computes the normalization constant of the density considered in \cite{said2017riemannianb}. In this case, we must use a different de Bruijn identity \cite{deBruijn}, which allows us to write 
\begin{equation}
    z_{\beta=4} (\sigma) = e^{\sigma^2 m \left( m - \frac{1}{2} \right)^2 } \underset{1 \le i, j \le 2m }{\mathrm{Pf}} \left[ 2 \langle x^{i-1} , x^{j-1} \rangle_4   \right] ,
\end{equation}
in terms of a Pfaffian with entries the skew-symmetric products given in \eqref{skewmombeta4}. Note that the Pfaffian is that of a $2m \times 2m $ skew-symmetric matrix. We plot the result as a function of $\sigma^2$ in Figure \ref{fig:z4vssigma}, and as a function of $m$ for various fixed values of $\sigma^2$ is Figure \ref{fig:z4vsrank}.\par
Moreover, we evaluate numerically $\log z_{\beta=4}  (\sigma) $ in Tables \ref{tab:zbeta4} and \ref{tab:zmbeta4}.\par

\par
\par

 \begin{small}
\begin{table}
\centering
\hspace*{-1cm}
\begin{tabular}{c ||c|c|c|c|c|c|c|c|c|c|c|c| }
$m ~ \vert ~ \sigma^2 $ & 0.1 & 0.2 & 0.3 & 0.4 & 0.5 & 0.6 & 0.7 & 0.8 & 0.9 & 1.0 & 1.1 & 1.2   \\
\hline
2 &  -5.460 & -4.008 & -3.115 & -2.454 & -1.917 & -1.460 & -1.057 & -0.693 & -0.359 & -0.04683 & 0.247 & 0.526  \\
3 & -12.993 & -8.527 & -5.737 & -3.633 & -1.902 & -0.403 & 0.940 & 2.171 & 3.318 & 4.402 & 5.435 & 6.430  \\
4 & -22.443 & -13.291 & -7.477 & -3.021 & 0.703 & 3.976 & 6.947 & 9.706 & 12.308 & 14.793 & 17.187 & 19.511  \\
5 & -33.163 & -17.533 & -7.439 & 0.4189 & 7.083 & 13.019 & 18.475 & 23.596 & 28.476 & 33.176 & 37.740 & 42.199  \\
6 & -44.620 & -20.597 & -4.830 & 7.629 & 18.342 & 28.002 & 36.977 & 45.480 & 53.649  & 61.572 & 69.319 & 76.931  \\
7 & -56.352 & -21.896 & 1.082 & 19.502 & 35.542 & 50.166 & 63.881 & 76.859 & 91.584 & 110.199 & 132.388 & 152.264 \\
  \hline \hline \\
$m ~ \vert ~ \sigma^2 $ & 1.3 & 1.4 &  1.5 &  1.6 & 1.7 & 1.8 & 1.9 & 2.0 & 2.1 & 2.2 & 2.4 & 2.4 \\
\hline 
2 &   0.794&   1.051 & 1.301 & 1.543 & 1.780& 2.011 & 2.239 & 2.462 & 2.683 & 2.901 & 3.116 & 3.330 \\
3 & 7.392 & 8.329 & 9.244 & 10.142 & 11.026 & 11.897 & 12.758 & 13.610 & 14.455 & 15.293 & 16.126 & 16.955 \\
4 &  21.779 & 24.002 & 26.189 & 28.346 & 30.478 & 32.591 & 34.686 & 36.767 & 38.837 & 40.896 & 42.947 & 44.991 \\
5 & 46.576 & 50.887 &   55.146 & 59.365 & 63.550 & 67.705 & 71.846 & 76.049 & 80.407 & 83.995 & 94.476 & 99.749 \\
6 & 83.662 & 94.386 &   108.756 & 120.963 & 137.087 & 134.410 & 165.131 & 180.287 & 171.118 & 179.404 & 230.606 & 211.766 \\
7 &  163.070 & 198.697 & 214.792  & 202.635 & 288.985 & 277.438 & 344.463 & 373.588 & 361.125 & 359.058 & 413.542 & 408.297 \\
\hline \hline 
\end{tabular}
\vspace{0.3cm}
\caption{$\log z_{\beta=4} (\sigma)$ for $m$ from $2$ to $7$ and $\sigma^2$ from $0.1$ to $2.4$. Evaluation time $0.109$s (whole table) on macbook.}
\label{tab:zbeta4}
\end{table}\par

\begin{table}[ht]
\centering
\begin{tabular}{c||c|c|c|c|c|c|c|c|c|c|}
$m $ & 1 & 2 & 3 & 4 & 5 & 6 & 7 & 8 & 9 & 10  \\
\hline
$- \log z_{4}$  &  0.693 & 8.789 & 23.061 & 42.750 & 67.319 & 92.443 & 117.980 & 141.695 & 167.451 & 191.576 \\
\hline \\
$m $ & 11 & 12 & 13 & 14 & 15 & 16 & 17 & 18 & 19 & 20  \\
\hline
$- \log  z_{4}$  & 211.572 & 235.962 & 256.665 & 275.833  & 293.453 & 308.834 & 321.895 & 334.330 & 343.452 & 356.662 \\
\hline 
\end{tabular}
\vspace{0.3cm}
\caption{$-\log z_{\beta=4} (\sigma)$ for $m$ from $1$ to $20$ and fixed $\sigma^2=0.02$. Evaluation time $0.110$s (whole table) on macbook.}
\label{tab:zmbeta4}
\end{table}
\end{small}
\par
\medskip
The determination of the density $\rho_{\beta=4} (r; \sigma)$ for arbitrary $m$ would entail constructing skew-orthogonal polynomials with respect to the skew-symmetric product $\langle \cdot, \cdot \rangle_4$, in analogy with the $\beta=1$ case. However, for the case $m=2$, it can be easily obtained from direct integration:
\begin{align}
    \rho_{\beta=4} (r; \sigma) \vert_{m=2} & = \frac{1}{z_{4} (\sigma) \vert_{m=2} } \int_{- \infty } ^{\infty} \frac{dr_2}{ \pi \sigma^2 } e^{- \frac{ c_4 (r ^2 + r_2^2 )}{\sigma^2}}  \left[ 2 \sinh \left( \frac{r - r_2}{2} \right) \right]^4 \notag \\
    & = \frac{1}{z_{4} (\sigma) \vert_{m=2} } \cdot  \frac{ e^{- \frac{c_{4} r^2 }{\sigma^2}} }{ \sqrt{\pi \sigma^2} } \int_{- \infty} ^{\infty} \frac{dr_2}{ \sqrt{\pi \sigma^2} } e^{- \frac{c_{4} r_2 ^2 }{\sigma^2}} \left[ 6 -4 \left( e^{r-r_2}  + e^{- r + r_2} \right) + \left( e^{2r-2r_2}  + e^{- 2r + 2r_2} \right)\right] \notag \\
    & = \frac{2}{z_{4} (\sigma) \vert_{m=2} } \cdot \frac{ e^{- \frac{c_{4} r^2 }{\sigma^2}} }{ \sqrt{c_{4} \pi \sigma^2}  } \left[ e^{ \sigma^2 /c_4} \cosh (2r) - 4 e^{\sigma^4 /(4 c_4)} \cosh (r) +3   \right] .
\end{align}
We recall from \eqref{normcbeta} that in our conventions $c_{\beta=4} =1$, but we have kept it explicitly here. The normalization $z_{\beta=4} (\sigma) $ at $m=2$ can be determined likewise, and we find 
\begin{equation}
    z_{\beta=4} (\sigma) \vert_{m=2} = \frac{e^{2 \sigma^2}}{2} - 2 e^{\sigma^2 /2 } + 3 .
\end{equation}

\section{Diffusion and kernel interpretation}
\label{sec:diffusion}

The first instance where the probability density \eqref{eq:gaussianpdf} appeared in physics is seemingly in the problem of vicious walkers in statistical physics \cite{fisher1984walks}. The formulation of the vicious random walker problem on a lattice, according to the so-called lock step model, consists in considering $m$ random walkers, each an even number of lattice spacings apart, on a one-dimensional lattice. Then, at regular time intervals, each walker must take either a step one lattice site to the left or a step one lattice site to the right, with equal probability 1/2. Coincidence of two walkers at the same site and time results in the annihilation of both walkers and hence the process ends (whence the name vicious). 

In the diffusive limit, this model describes a set of non-intersecting Brownian motions and one obtains the Gaussian Riemannian density \eqref{eq:gaussianpdf} and also the ones corresponding to other symmetric spaces, as we shall see, by consideration of the classical, determinantal formalism of non-intersecting random walkers and non-intersecting Brownian motion, initiated by Karlin and McGregor in the seminal work \cite{karlin1959coincidence} (see more recent work and textbooks \cite{grabiner1999brownian,katori2016bessel,BaikDeiftSuidanBook} for a clear description, including a modern explanation of Dyson's Brownian motion \cite{dyson1962brownian}).

Given the relevance of diffusion processes in the statistical analysis of data \cite{coifman2006diffusion,botev2010kernel}, it is worth to dwell further into the diffusive origins and interpretations of the probability densities discussed above.

In \cite{fisher1984walks}, the probability density that all walkers survive at time $t$ is obtained, giving an expression in terms of the positions $(x_1, \dots, x_m)$ of the form

\begin{equation}
   c_t (x )=\frac{1}{(2\pi t)^{m/2}}
	\exp\left(\frac{-|x|^2}{2t}\right)
	\prod_{1 \le i < j \le m }[\exp(x_j/t)-\exp(x_i/t)],
	\label{Qx}
\end{equation}
which, again using r.h.s of \eqref{change}, is the probability density function of \eqref{ZCS}, up to a shift of variables. Therefore, this density is the solution of a diffusion process and the density satisfies a heat equation \cite{Forrester:Vicious}.

Interestingly, the problem of $m$ vicious walkers on a one-dimensional lattice has an equivalent description in terms of Brownian motion on the Weyl chamber of a Lie group \cite{grabiner1999brownian,deHaroTierz:04}. All the densities discussed in this work emerge in a natural way in such a context. To see this, we quote a basic theorem \cite{grabiner1999brownian} which states that if $c_t$ is the density function for a continuous stochastic
process, then for absorbing boundary conditions, it holds
\begin{equation}
  b_{t} (x, \eta ) =\sumweyl\mathrm{sgn}(w) c_t(w(x)-\eta) . \label{brownabs}
\end{equation}
In this formula, $\eta=(\eta_1, \dots, \eta_m)$ is the vector of initial positions of the $m$ particles, $b_t (x, \eta)$ is the probability density for the particles to be at $x$ given the initial positions $\eta$, and the sum is over elements $w$ of the Weyl group $W$. 

These expressions are all determinantal and this theorem applies to standard Brownian motion, in the Weyl
chambers of $A_{m-1}$, $B_m$ (and thus $C_m$), and $D_m$, with absorbing or reflecting boundary conditions. The measure for unconstrained standard Brownian motion is $c_t( x )=\prod_{i=1} ^{m} N_t(x_i)$, where $N_t$ is the
normal distribution function with mean 0 and variance $t$.  

Consider the root system $A_{m-1}$, with associated Weyl group the symmetric group $S_m$. The principal Weyl chamber is characterized by $x_1>x_2>\cdots>x_m$. This models the Brownian motion of $m$ independent particles in one dimension. With absorbing boundary
conditions, collisions are forbidden and the sum can be written 
as a determinant, which gives
\begin{equation}
  b_{t} (x, \eta ) =\det_{1 \le i,j \le m }\left[  N_t(x_i-\eta_j) \right].
\label{anbrown}
\end{equation}
This determinant gives the probability for $n$ particles that start at positions $(\eta_1, \dots, \eta_m)$ and are in independent Brownian motion to be at positions $(x_1, \dots, x_m)$ at time $t$ without having collided.\par
The expression \eqref{anbrown} can be written as
\begin{equation}
  b_{t} (x, \eta ) =\frac{1}{(2\pi t)^{m/2}}
	\exp\left(\frac{-|x|^2-|\eta|^2}{2t}\right)
	\det_{1 \le i,j \le m }\left[ 
            \exp\left(\frac{x_i\eta_j}{t}\right)\right].
  \label{expvand}
\end{equation}
The $\eta_j$ are all integers and, if in addition we choose $ \eta_j=m-j $, the determinant on the right-hand side is the Vandermonde determinant in the variables $e^{x_i /t}$, equal to
\begin{equation}
  \prod_{1 \le i < j \le m } \left( e^{x_j /t} - e^{x_i /t}  \right)  .\label{anvand}
\end{equation}
Therefore, we have that, for this equal spacing condition on $\eta_j$:
\begin{equation}
  b_{t} (x, \eta )= \exp\left(\frac{-|\eta|^2}{2t}\right)c_t (x )=\exp\left(\frac{-(m-1)(2m-1)/6}{2t}\right)c_t (x ),
\end{equation}
where $c_t (x )$ is \eqref{Qx}. Therefore, because of \eqref{change}, the density is that of a Gaussian distribution on the Riemannian space of real symmetric matrices introduced in \cite{said2017riemannian,cheng2013novel,said2017riemannianb}, and studied here in Section \ref{sec:beta-14}. 

The corresponding Gaussian density for the space of Hermitian matrices appears from the above when studying the probability of reunion, using for example the extensivity property of probabilities \cite[Eq. (14)]{deHaroTierz:04}, which in turn is related to the reproducing property of the (determinantal) kernel \eqref{expvand}.

For a more general setting involving any non-negative integers $\eta_j$, the determinant above is the product of this Vandermonde determinant and the Schur function \cite{ForresterBook}
\begin{equation}
  s_{\eta_1-m+1,\eta_2-m+2,\ldots,\eta_m}  \left(  e^{x_1 /t} , \dots, e^{x_m /t} \right) ,
\end{equation}
giving a more general probability density than the ones discussed here, but relevant (and computable), in other contexts \cite{deHaroTierz:04,TierzDolivet,tierz2010schur,giasemidis2014torus}. See also \cite{baik2007random} to see how the Stieltjes-Wigert ensemble emerges directly from the Karlin-McGregor formula when considering Brownian bridges.

For root system $B_m$, the corresponding Weyl group is the hyperoctahedral group, which includes permutations with any number of sign changes. Its Weyl chamber is $x_1>x_2>\cdots>x_m>0$.
This also models $m$ independent particles in one dimension, with an
additional wall at $x=0$ \cite{grabiner1999brownian}. In the Brownian motion setting at hand, the formalism gives 
\begin{equation}
   b_{t} (x, \eta )  =\frac{1}{(2\pi t)^{m/2}}
	\exp\left(\frac{-|x|^2-|\eta|^2}{2t}\right)
	\det_{1 \le i,j \le m }\left[  
        \exp\left(\frac{x_i\eta_j}{t}\right)
        -\exp\left(\frac{-x_i\eta_j}{t}\right)\right].
  \label{bnexp}
\end{equation}
This determinant gives the measure for $m$ particles which start at $(\eta_1, \dots, \eta_m)$ to be at $(x_1, \dots, x_m)$ at time $t$, neither having collided nor having touched $x=0$. Choosing  again $\eta_j=m-j$, the determinant in \eqref{bnexp} can be written in a Vandermonde form \cite{grabiner1999brownian}. Doing so and with some rewriting, that part gives 
\begin{equation}
    \prod_{1 \le i < j \le m } 4 \sinh \left( \frac{\vert \lambda_i - \lambda_j \vert}{2}  \right) \sinh \left( \frac{\vert \lambda_i + \lambda_j \vert}{2}  \right) ~ \prod_{i=1} ^{m}  2 \sinh \left( \left\lvert \frac{\lambda_i}{2} \right\rvert \right) . 
\end{equation}
This expression is associated with the Jacobian for matrix integration on $\mathfrak{o} (2m+1)$.\par
These non-intersecting Brownian motions are interwoven with many other areas. One of this relationships is the connection with the Harish-Chandra-Itzykson-Zuber integral \cite{katori2002scaling,katori2004symmetry,katori2016bessel}. For $\mathfrak{o} (m)$ and $\mathfrak{sp} (2m)$, this has been further developed in \cite{forrester2019orthogonal} where, in addition, it is proven that the density $b_t (x, \eta) $ satisfies a diffusion equation.\par

Likewise, for $D_m$, with Weyl group the even hyperoctahedral group, which includes permutations with an even number of sign changes, the principal Weyl chamber is
$x_1>x_2>\cdots>x_m$, with $x_{m-1}>-x_m$. This does not give a natural model for $m$ particles in one dimension, but one can still find a determinantal expression leading to the $\mathfrak{so} (2m)$ case studied in Appendix \ref{app:SoandSp}, see Eq. \eqref{so2n}, and also appeared in statistical work in \cite{said2017gaussian} (but not analyzed analytically there). In Appendix \ref{app:SoandSp} we study that case and the one corresponding to $\mathfrak{sp} (2m)$, Eq. \eqref{sp2n}, which is closely related to the odd orthogonal case quoted above (see also \cite{forrester2019orthogonal}).\par
\medskip
It is worth mentioning that $b_t (x, \eta) $ is itself a reproducing kernel, as can be quickly checked (again, see for example \cite[Eq. (14)]{deHaroTierz:04}), even though this fundamental property does not seem to have been exploited, in spite of the broad literature on the subject.\par
It would be interesting if this fact could be used in a statistical context, either in conjunction with the results described and cited or on its own, given the very well established relevance of reproducing kernels in traditional statistical analysis \cite{hendriks1990nonparametric} and, at the same time, the remarkable and deep relevance of diffusion, in the statistical analysis of data, at many different levels \cite{coifman2006diffusion,botev2010kernel}.

It should be stressed that this is a different reproducing kernel from the one that immediately emerges by identifying the probability density in terms of a random matrix ensemble \eqref{ZSW}. Random matrix theory \cite{MehtaBook} associates to \eqref{ZSW} a reproducing kernel of the form:
\begin{equation}
K_{m}\left( x,y\right) = \sqrt{w (x) w (y)} ~ \sum_{k=0}^{m-1}p_{k}\left( x\right) p_{k}(y).
\end{equation}
where $w (x)$ is the Stieltjes-Wigert weight and $p_{k}\left( x\right)$ are the Stieltjes-Wigert polynomials, introduced in Section \ref{sec:beta2}. This is a delta sequence type of kernel \cite{hendriks1990nonparametric}. Since the polynomials are $q$-deformations of Hermite polynomials, the kernel is a one-parameter deformation of a Hermite kernel, used in probability density estimations \cite{walter1977properties,hendriks1990nonparametric}. From it, in principle any marginal of the probability density can be obtained. The result in Section \ref{sec:rhobeta2} is an example of this, corresponding to the diagonal limit of the kernel. See \cite{Forrester:crystal} for explicit evaluations, and \cite{Forrester:new} for recent developments along these lines.\par
\medskip
An interesting open problem is to construct the skew-orthogonal polynomials associated to the Stieltjes-Wigert (log-normal) weight function. This would put on equal footing the analytical results for the real-symmetric and quaternionic models with the full solution of the case of Hermitian matrices. That would be a new result from the random matrix theory point of view as well, since only classical orthogonal polynomials \cite{AFNvM,AHvM:Pfaff} and, more recently, semi-classical polynomials \cite{mays2020tracy} have been studied, whereas skew-polynomials in the $q$-deformed setting, to which the Stieltjes-Wigert polynomials belong, remain unstudied.

\vspace{0.5cm}
\subsection*{Acknowledgments}

We thank Professors Peter Forrester and Gregory Schehr for correspondence. The work is partially supported by FCT Project PTDC/MAT-PUR/30234/2017. The work of LS is supported by the FCT through the doctoral grant SFRH/BD/129405/2017.

\appendix

\section{Skew-orthogonal polynomials}
\label{app:skewpol}

This Appendix collects the explicit expressions of the skew-orthogonal polynomials and the corresponding $\widetilde{\phi} _n (x)$, used in Section \ref{sec:rhobeta1} in the computation of the density of states for the case of real symmetric matrices ($\beta=1$). The first few skew-orthogonal polynomials are 
\begin{align}
    \widetilde{p}_0 (x) & = 1, \\ \widetilde{p}_1 (x) & = x ,  \notag \\
    \widetilde{p}_2 (x) & = x^2 - x e^{\frac{5}{2} \sigma^2 } \frac{ \mathrm{erf} (\vert \sigma \vert ) }{\mathrm{erf} (\vert \sigma \vert /2 ) } + e^{4 \sigma^2}  , \notag \\
    \widetilde{p}_3 (x) & = x^3 + x^2  \frac{ \left(-e^{\frac{7
   \sigma ^2}{4}} \text{erf}\left(\frac{ \vert \sigma \vert }{2}\right)-e^{\frac{11 \sigma ^2}{2}}
   \text{erf}\left(\vert \sigma \vert \right)+e^{\frac{19 \sigma ^2}{4}} \text{erf}\left(\frac{3 \vert \sigma
   \vert }{2}\right)\right)}{\left(e^{2 \sigma ^2}+1\right) \text{erf}\left(\frac{\vert \sigma
   \vert }{2}\right)-e^{\frac{5 \sigma ^2}{4}} \text{erf}\left(\vert \sigma \vert \right)}  \notag \\
   & + x \frac{\left(e^{8 \sigma ^2} \text{erf}\left(\frac{\vert \sigma \vert }{2}\right)+e^{\frac{17 \sigma
   ^2}{4}} \text{erf}\left(\vert \sigma \vert \right)-e^{6 \sigma ^2} \text{erf}\left(\frac{3 \vert \sigma \vert }{2}\right)\right)}{\left(e^{2 \sigma ^2}+1\right) \text{erf}\left(\frac{\vert \sigma
   \vert }{2}\right)-e^{\frac{5 \sigma ^2}{4}} \text{erf}\left( \vert \sigma \vert \right)}
   + \frac{\left(-e^{\frac{23 \sigma ^2}{4}}-e^{\frac{35 \sigma ^2}{4}}\right) \text{erf}\left(\frac{ \vert \sigma \vert }{2}\right)+e^{\frac{15 \sigma ^2}{2}} \text{erf}\left( \vert \sigma \vert \right)}{\left(e^{2 \sigma ^2}+1\right)
   \text{erf}\left(\frac{\vert  \sigma \vert }{2}\right)-e^{\frac{5 \sigma ^2}{4}} \text{erf}\left(\vert \sigma
   \vert \right)}    \notag 
\end{align}
and the associated $\widetilde{\phi}$ functions, in the variable $x= e^{u}$, are 
\begin{align}
    \sqrt{2} \widetilde{\phi}_0 (e^{u} ) & = e^{\frac{1}{2} \sigma^2} \mathrm{erf} \left( \frac{u - \sigma^2}{\sqrt{2 \sigma^2} } \right), \quad \sqrt{2} \widetilde{\phi}_1 (e^{u}) = e^{2\sigma^2} \mathrm{erf} \left( \frac{u - 2 \sigma^2}{\sqrt{2 \sigma^2} } \right) ,  \\
     \sqrt{2} \widetilde{\phi}_2 (e^{u}) & = e^{\frac{9}{2} \sigma^2} \left[  \mathrm{erf} \left( \frac{u - 3\sigma^2}{\sqrt{2 \sigma^2} } \right) - \mathrm{erf} \left( \frac{u - 2\sigma^2}{\sqrt{2 \sigma^2} } \right) + \mathrm{erf} \left( \frac{u - \sigma^2}{\sqrt{2 \sigma^2} } \right) \right] , \notag \\
      \sqrt{2} \widetilde{\phi}_3 (e^{u} ) & =
     e^{8 \sigma ^2} \text{erf}\left(\frac{u-4 \sigma ^2}{\sqrt{2} \sqrt{\sigma ^2}}\right)+   e^{\frac{9
   \sigma ^2}{2}}\text{erf}\left(\frac{u-3 \sigma ^2}{ \sqrt{2 \sigma ^2}}\right)  \left[   \frac{ \left(-e^{\frac{7
   \sigma ^2}{4}} \text{erf}\left(\frac{ \vert \sigma \vert }{2}\right)-e^{\frac{11 \sigma ^2}{2}}
   \text{erf}\left(\vert \sigma \vert \right)+e^{\frac{19 \sigma ^2}{4}} \text{erf}\left(\frac{3 \vert \sigma
   \vert }{2}\right)\right)}{\left(e^{2 \sigma ^2}+1\right) \text{erf}\left(\frac{\vert \sigma
   \vert }{2}\right)-e^{\frac{5 \sigma ^2}{4}} \text{erf}\left(\vert \sigma \vert \right)}     \right]  \notag  \\
   & +e^{2 \sigma ^2} \text{erf}\left(\frac{u-2
   \sigma ^2}{\sqrt{2 \sigma ^2}}\right)  \left[  \frac{\left(e^{8 \sigma ^2} \text{erf}\left(\frac{\vert \sigma \vert }{2}\right)+e^{\frac{17 \sigma
   ^2}{4}} \text{erf}\left(\vert \sigma \vert \right)-e^{6 \sigma ^2} \text{erf}\left(\frac{3 \vert \sigma \vert }{2}\right)\right)}{\left(e^{2 \sigma ^2}+1\right) \text{erf}\left(\frac{\vert \sigma
   \vert }{2}\right)-e^{\frac{5 \sigma ^2}{4}} \text{erf}\left( \vert \sigma \vert \right)}   \right]   \notag \\
   & +e^{\frac{\sigma ^2}{2} }
   \text{erf}\left(\frac{u-\sigma ^2}{ \sqrt{ 2 \sigma ^2}}\right) \left[ \frac{\left(-e^{\frac{23 \sigma ^2}{4}}-e^{\frac{35 \sigma ^2}{4}}\right) \text{erf}\left(\frac{ \vert \sigma \vert }{2}\right)+e^{\frac{15 \sigma ^2}{2}} \text{erf}\left( \vert \sigma \vert \right)}{\left(e^{2 \sigma ^2}+1\right)
   \text{erf}\left(\frac{\vert  \sigma \vert }{2}\right)-e^{\frac{5 \sigma ^2}{4}} \text{erf}\left(\vert \sigma
   \vert \right)}  \right] \notag .
\end{align}\par

\section{The case of other symmetric spaces}
\label{app:SoandSp}

The methods presented in the main text can be extended to other symmetric spaces. As discussed previously, switching to other spaces in the Cartan classification corresponds, on the random matrix theory side, to consider matrix integrals with different integration domain. We restrict ourselves to hyperbolic spaces and $\beta=1$ for concreteness.\par

We consider the partition function of the $\beta=1$ ensemble of $\mathfrak{so}(2m)$ matrices. The choice $\mathfrak{so}$ belongs to the $D$-class in the Cartan classification, and $\beta=1$ (orthogonal symmetry) means that we are working in the tangent space at the origin to the coset space $SO(2m) / U(m)$, hence in the DIII class \cite{ZirnbauerReview}. The model also appears in the latter part of \cite{said2017gaussian} where it is shown to be related to integration on a Siegel disc.

The partition function is
\begin{equation}
     z_{\beta=1} ^{\mathfrak{so}(2m)} (\sigma) = \frac{1}{(2 m)!} \int_{\R^m} \prod_{1 \le i < j \le m } 4 \sinh \left( \frac{\vert r_i - r_j \vert}{2}  \right) \sinh \left( \frac{\vert r_i + r_j \vert}{2}  \right) ~ \prod_{i=1} ^{m}   e^{- \frac{r_i ^2 }{2 \sigma^2}  } \frac{ dr_i }{\sqrt{\pi \sigma^2} } . 
     \label{so2n}
\end{equation}
We are now working with $2m \times 2m$ matrices, therefore the Weyl group permuting the $2m$ eigenvalues is of order $(2m)!$. We then use the invariance of the integral under such permutations to restrict the integration domain to the principal Weyl chamber 
\begin{equation}
    r_1 \ge r_2 \ge \cdots \ge r_m \ge 0 
\end{equation}
and obtain 
\begin{equation}
     z_{1} ^{\mathfrak{so}(2m)} (\sigma) = \int_{0 \le r_m \le \cdots \le r_1 < \infty } \prod_{1 \le i < j \le m } 4 \sinh \left( \frac{r_i - r_j }{2}  \right) \sinh \left( \frac{ r_i + r_j }{2}  \right) ~ \prod_{i=1} ^{m}   e^{- \frac{r_i ^2 }{2 \sigma^2}  } \frac{ dr_i }{\sqrt{\pi \sigma^2} } . 
\end{equation}
Using the simple property 
\begin{equation}
    4 \sinh \left( \frac{r_i - r_j }{2}  \right) \sinh \left( \frac{ r_i + r_j }{2}  \right) = 2 \left[ \cosh r_i - \cosh r_j \right] 
\end{equation}
and changing variables $x_i = \cosh r_i$, we arrive at 
\begin{equation}
\label{eq:z1sochangevar}
     z_{1} ^{\mathfrak{so}(2m)} (\sigma) = 2^{m (m-1)/2} \int_{1 \le x_m \le \cdots \le x_1 < \infty } \prod_{1 \le i < j \le m } (x_i - x_j ) ~ \prod_{i=1} ^{m}   w^{\mathfrak{so}} (x; \sigma) \frac{ dx_i }{\sqrt{\pi \sigma^2} } ,
\end{equation}
where we have defined 
\begin{equation}
    w^{\mathfrak{so}} (x; \sigma) = \frac{ \exp \left\{ - \frac{1}{2 \sigma^2 } \left[ \log \left( x + \sqrt{x ^2 -1} \right) \right]^2 \right\}  }{  \log \left( x + \sqrt{x ^2 + 1} \right) } .
\end{equation}
The function $ w^{\mathfrak{so}} $ satisfies 
\begin{equation}
    w^{\mathfrak{so}} (x; \sigma) dx \vert_{x = \cosh r }=  e^{- \frac{r^2}{2 \sigma^2}} dr .
\end{equation}
The expression \eqref{eq:z1sochangevar} is now suitably written to apply the de Bruijn identity \cite{deBruijn}. To this aim, we introduce the symmetric inner product 
\begin{align}
    \left( f, g \right)_2 ^{\mathfrak{so}} & = \int_1 ^{\infty} \frac{dx}{\sqrt{\pi \sigma^2}}  w^{\mathfrak{so}} (x; \sigma/\sqrt{2} )  f(x) g (x) \notag \\
    & = \int_0 ^{\infty} \frac{dr}{\sqrt{\pi \sigma^2}} e^{- \frac{r^2}{\sigma^2}} f ( \cosh (r)) g ( \cosh (r))  = \frac{1}{2} \int_{- \infty} ^{\infty} \frac{dr}{\sqrt{\pi \sigma^2}} e^{- \frac{r^2}{\sigma^2}} f ( \cosh (r)) g ( \cosh (r)) .
\end{align}
We will need 
\begin{align}
    \left( 1, x^{j-1} \right)_2 ^{\mathfrak{so}}  & = \frac{1}{2} \int_{- \infty} ^{\infty} \frac{dr}{\sqrt{\pi \sigma^2}} e^{- \frac{r^2}{\sigma^2}} ( \cosh r )^{j-1} \notag \\
    & = \frac{1}{2^j} \sum_{\ell=0} ^{j-1} \left( \begin{matrix} j-1 \\ \ell  \end{matrix} \right) \int_{- \infty} ^{\infty} \frac{dr}{\sqrt{\pi \sigma^2}} e^{- \frac{r^2}{\sigma^2}}  e^{r (j-1 - 2 \ell )} \notag \\
    & = \frac{1}{2^j} \sum_{\ell=0} ^{j-1} \left( \begin{matrix} j-1 \\ \ell  \end{matrix} \right) e^{\frac{\sigma^2}{2} (j-1 - 2 \ell)^2 } .
\end{align}
Moreover, we introduce the skew-symmetric product 
\begin{align}
    \left\langle f, g \right\rangle_1 ^{\mathfrak{so}} & = \frac{1}{2} \int_1 ^{\infty} \frac{dx}{\sqrt{\pi \sigma^2}}  w^{\mathfrak{so}} (x; \sigma)  f(x) \int_1 ^{\infty} \frac{dy}{\sqrt{\pi \sigma^2}}  w^{\mathfrak{so}} (y; \sigma)   g (y) ~ \mathrm{sign} (x-y) \notag \\
    & = \int_0 ^{\infty} \frac{dr_1}{\sqrt{\pi \sigma^2}} e^{- \frac{r_1 ^2}{2 \sigma^2}} f ( \cosh (r_1)) \int_0 ^{\infty} \frac{dr_2}{\sqrt{\pi \sigma^2}} e^{- \frac{r_2 ^2}{2 \sigma^2}} g ( \cosh (r_2))  \frac{\mathrm{sign} (r_1 - r_2)}{2} ,
\end{align}
which for monomials reads 
\begin{align}
    \left\langle x^{i-1}, x^{j-1} \right\rangle_1 ^{\mathfrak{so}} & = \int_0 ^{\infty} \frac{dr_1}{\sqrt{\pi \sigma^2}} e^{- \frac{r_1 ^2}{2 \sigma^2}}  ( \cosh r_1)^{i-1} \int_0 ^{\infty} \frac{dr_2}{\sqrt{\pi \sigma^2}} e^{- \frac{r_2 ^2}{2 \sigma^2}} ( \cosh r_2 )^{j-1}  \frac{\mathrm{sign} (r_1 - r_2)}{2} \notag \\
    & = \frac{1}{2^{i+j-2}} \sum_{k=0} ^{i-1} \left( \begin{matrix} i-1 \\ k  \end{matrix} \right)  \sum_{\ell=0} ^{j-1} \left( \begin{matrix} j-1 \\ \ell  \end{matrix} \right) \notag \\
    & \quad \times \int_{0} ^{\infty} \frac{dr_1}{\sqrt{\pi \sigma^2}} e^{- \frac{r_1 ^2}{2 \sigma^2}}  e^{ r_1 (i-1 - 2k) } \int_{0} ^{\infty} \frac{dr_2}{\sqrt{\pi \sigma^2}} e^{- \frac{r_2 ^2}{2 \sigma^2}}  e^{ r_2 (j-1 - 2\ell ) }  \frac{\mathrm{sign} (r_1 - r_2)}{2} \notag \\
    & = \frac{1}{2^{i+j-1}} \sum_{k=0} ^{i-1} \left( \begin{matrix} i-1 \\ k  \end{matrix} \right)  \sum_{\ell=0} ^{j-1} \left( \begin{matrix} j-1 \\ \ell  \end{matrix} \right) e^{\frac{\sigma^2}{2} (j-1 - 2 \ell)^2 } \notag \\
     & \quad \times  \int_{0} ^{\infty} \frac{dr_1}{\sqrt{2 \pi \sigma^2}} e^{- \frac{r_1 ^2}{2 \sigma^2} + r_1 (i-1 - 2k) }  \left[ \mathrm{erf} \left( \frac{ r_1 + \sigma^2 (j-1 - 2 \ell ) }{\sqrt{2 \sigma^2}} \right) + \mathrm{erf} \left( \frac{ r_1 - \sigma^2 (j-1 - 2 \ell ) }{\sqrt{2 \sigma^2}} \right)   - 1 \right] .
\end{align}
We are now ready to apply the de Bruijn identity to \eqref{eq:z1sochangevar}. We obtain the Pfaffian form:
\begin{equation}
\label{eq:z1soPfaffeven}
    z_1 ^{\mathfrak{so}(2m)} (\sigma) \vert_{m \text{ even}}=  2^{m (m-1)/2} \underset{1 \le i,j \le m}{\mathrm{Pf}} \left[ 2 \langle x^{i-1}, x^{j-1} \rangle_1 ^{\mathfrak{s0}}  \right] 
\end{equation}
for even $m$, and 
\begin{equation}
\label{eq:z1soPfaffodd}
     z_1 ^{\mathfrak{so}(2m)} (\sigma) \vert_{m \text{ odd}}=  2^{m (m-1)/2}  \underset{1 \le i,j \le m+1}{\mathrm{Pf}} \left[ \begin{matrix}  \left[  2 \langle x^{i-1}, x^{j-1} \rangle_1 ^{\mathfrak{so}} \right]_{i,j \le m} & \left[ 2 \left( 1, x^{i-1} \right)_2 ^{\mathfrak{so}} \right]_{i \le m} \\ \left[ - 2 \left( 1, x^{j-1} \right)_2 ^{\mathfrak{so}} \right]_{j \le m}  & 0 \end{matrix}  \right] 
\end{equation}
for odd $m$.\par
The procedure extends to the $\mathfrak{sp} (2m)$ case. For $\beta=1$, this corresponds to the class CI in the Cartan classification \cite{ZirnbauerReview}. The partition function is \cite{ForresterBook} 
\begin{equation}
    z_1 ^{\mathfrak{sp}(2m)} (\sigma) = \frac{1}{(2 m)!} \int_{\R^m} \prod_{1 \le i < j \le m } 4 \sinh \left( \frac{\vert r_i - r_j \vert}{2}  \right) \sinh \left( \frac{\vert r_i + r_j \vert}{2}  \right) ~ \prod_{i=1} ^{m}  2 \sinh (\vert r_i \vert ) ~ e^{- \frac{r_i ^2 }{2 \sigma^2}  } \frac{ dr_i }{\sqrt{\pi \sigma^2} } . 
    \label{sp2n}
\end{equation}
We mimic the steps above, using the invariance to restrict the integration domain to the principal Weyl chamber and change variables $x_i = \cosh r_i$. We arrive at 
\begin{equation}
    z_1 ^{\mathfrak{sp}(2m)} (\sigma) =  2^{m (m+1)/2} \int_{ 1 \le x_m \le \cdots \le x_1 < \infty } \prod_{1 \le i < j \le m } (x_i - x_j) ~ \prod_{i=1} ^{m}   ~ e^{- \frac{1 }{2 \sigma^2}  \left[ \log \left( x_i + \sqrt{x_i ^2 -1} \right) \right]^2 } \frac{ dx_i }{\sqrt{\pi \sigma^2} } . 
\end{equation}
The de Bruijn identity \cite{deBruijn} gives 
\begin{equation}
\label{eq:z1spPfaffeven}
    z_1 ^{\mathfrak{sp}(2m)} (\sigma) \vert_{m \text{ even}}=  2^{m (m+1)/2} \underset{1 \le i,j \le m}{\mathrm{Pf}} \left[ 2 \langle x^{i-1}, x^{j-1} \rangle_1 ^{\mathfrak{sp}}  \right] 
\end{equation}
for even $m$, and 
\begin{equation}
\label{eq:z1spPfaffodd}
     z_1 ^{\mathfrak{sp}(2m)} (\sigma) \vert_{m \text{ odd}}=  2^{m (m+1)/2}  \underset{1 \le i,j \le m+1}{\mathrm{Pf}} \left[ \begin{matrix}  \left[  2 \langle x^{i-1}, x^{j-1} \rangle_1 ^{\mathfrak{sp}} \right]_{i,j \le m} & \left[ 2 \left( 1, x^{i-1} \right)_2 ^{\mathfrak{sp}} \right]_{i \le m} \\ \left[ - 2 \left( 1, x^{j-1} \right)_2 ^{\mathfrak{sp}} \right]_{j \le m}  & 0 \end{matrix}  \right] 
\end{equation}
for odd $m$. The inner product $\left( \cdot , \cdot \right)_2 ^{\mathfrak{sp}} $ in \eqref{eq:z1spPfaffodd} is given by \begin{align}
    \left( 1, x^{j-1} \right)_2 ^{\mathfrak{sp}} & = \int_0 ^{\infty} \frac{ dr }{\sqrt{\pi \sigma^2} } e^{ - r^2 / \sigma^2 } \sinh (r) [\cosh (r) ]^{j-1} \notag \\
    & = 2^{-j} \int_0 ^{\infty} \frac{ dr }{\sqrt{\pi \sigma^2} } e^{ - r^2 / \sigma^2 } \sum_{\ell=0} ^{j-1} \left( \begin{matrix} j-1 \\ \ell \end{matrix} \right)\left( e^{r(j-2 \ell ) }  -  e^{r(j-2 \ell -2) } \right) \notag \\
    & = \frac{1}{2^{j+1}} \sum_{\ell=0} ^{j-1} \left( \begin{matrix} j-1 \\ \ell \end{matrix} \right) \left[ e^{(j - 2 \ell)^2 \frac{\sigma^2}{4}  } \left( 1 + \mathrm{erf} \left( \frac{ \sigma (j - 2 \ell )}{2} \right)  \right) -   e^{(j - 2 \ell -2 )^2 \frac{\sigma^2}{4}  } \left( 1 + \mathrm{erf} \left( \frac{ \sigma (j - 2 \ell -2 )}{2} \right)  \right) \right] .
\end{align}
and the skew-symmetric product $\langle \cdot , \cdot \rangle_2 ^{\mathfrak{sp}} $ for monomials is given by  
\begin{align}
    \langle x^{i-1}, x^{j-1} \rangle_1 ^{\mathfrak{sp}} & = \frac{1}{2} \int_{0} ^{\infty}  \frac{ dr_1 }{\sqrt{\pi \sigma^2} } e^{ - \frac{r_1 ^2 }{2 \sigma^2}  } [\cosh (r_1)]^{i-1} \sinh (r_1) \int_{0} ^{\infty}  \frac{ dr_2 }{\sqrt{\pi \sigma^2} } e^{ - \frac{r_2^2 }{2 \sigma^2} } [\cosh (r_2)]^{j-1} \sinh (r_2) ~ \mathrm{sign} (r_1-r_2) \notag  \\
    & = 2^{-i-j} \sum_{k=0} ^{i-1} \sum_{\ell=0}^{j-1}  \left( \begin{matrix} i-1 \\ k \end{matrix} \right) \left( \begin{matrix} j-1 \\ \ell \end{matrix} \right) \int_0 ^{\infty} \frac{ dr_1 }{\sqrt{\pi \sigma^2} } e^{ - \frac{r_1 ^2 }{2 \sigma^2}  }  \left( e^{r_1(i-k)} - e^{r_1(i-k-2)} \right)\notag \\
    & \hspace{3cm} \times  \int_0 ^{\infty} \frac{ dr_2 }{\sqrt{\pi \sigma^2} } e^{ - \frac{r_2 ^2 }{2 \sigma^2}  }  \left( e^{r_2(j - \ell )} - e^{r_(j - \ell -2)} \right) \frac{ \mathrm{sign} (r_1-r_2)}{2} \notag \\
    & = \frac{1}{2^{i+j+1}} \sum_{k=0} ^{i-1} \sum_{\ell=0}^{j-1}  \left( \begin{matrix} i-1 \\ k \end{matrix} \right) \left( \begin{matrix} j-1 \\ \ell \end{matrix} \right) \int_0 ^{\infty} \frac{dr_1}{\sqrt{\pi \sigma^2}} e^{ - \frac{r_1 ^2 }{2 \sigma^2}  }  \left( e^{r_1(i-k)} - e^{r_1(i-k-2)} \right) \notag \\
    & \hspace{3cm} \times \left\{ e^{(j - 2 \ell)^2 \frac{\sigma^2}{4} } \left[ \mathrm{erf} \left( \frac{ \sigma (j - 2 \ell )}{2} \right) -1 -2 \mathrm{erf} \left( \frac{ \sigma (j - 2 \ell )}{2} - \frac{r_1}{\sigma} \right) \right] \right. \notag \\ & \hspace{3cm} \left.  -   e^{(j - 2 \ell -2 )^2 \frac{\sigma^2}{4} } \left[  \mathrm{erf} \left( \frac{ \sigma (j - 2 \ell -2 )}{2}  \right) -1 -2 \mathrm{erf} \left( \frac{ \sigma (j - 2 \ell -2 )}{2}  - \frac{r_1}{\sigma} \right) \right] \right\} .
\end{align}

We note that these models, in the trigonometric version and with squared interaction terms (that is, $\beta=2$) have been solved fully in \cite{garcia2020matrix}, using the connection to determinants, instead of Pfaffians, of Toeplitz+Hankel matrices. One method that could be applied here would be for example the use of expansions in Schur polynomials, see \cite[Theorem 4]{garcia2020matrix} and the proof therein.

\bibliography{SW_matrix_biblio.bib}

\providecommand{\href}[2]{#2}\begingroup\raggedright\begin{thebibliography}{10}

\bibitem{helgason}
S.~Helgason, \emph{{Differential geometry, Lie groups, and symmetric spaces}},
  vol.~34 of \emph{Graduate Studies in Mathematics}. American Mathematical
  Society, Providence, RI, 2001.

\bibitem{terras2}
A.~Terras, \emph{{Harmonic analysis on symmetric spaces and applications II}}.
  Springer-Verlag, New York, NY, 1988,
  \href{https://doi.org/10.1007/978-1-4612-3820-1}{10.1007/978-1-4612-3820-1}.

\bibitem{bhatia}
R.~Bhatia, \emph{Positive definite matrices}, Princeton Series in Applied
  Mathematics. Princeton University Press, 41 William Street, Princeton, NJ,
  2007.

\bibitem{statistics}
L.~T. Skovgaard, \emph{{A Riemannian Geometry of the Multivariate Normal
  Model}}, {\emph{Scandinavian J. Stat.} {\bfseries 11} (1984) 211}
\newblock Available \href{http://www.jstor.org/stable/4615960}{online}.

\bibitem{atkinson}
C.~Atkinson and A.~F.~S. Mitchell, \emph{{Rao's Distance Measure}},
  {\emph{Indian J. Stat.} {\bfseries 43} (1981) 345}
\newblock Available \href{http://www.jstor.org/stable/25050283}{online}.

\bibitem{pennec3}
V.~Arsigny, P.~Fillard, X.~Pennec and N.~Ayache, \emph{{Log-Euclidean metrics
  for fast and simple calculus on diffusion tensors}},
  \href{https://doi.org/10.1002/mrm.20965}{\emph{Magnetic Resonance in
  Medicine} {\bfseries 56} (2006) 411}.

\bibitem{pennec2}
X.~Pennec, P.~Fillard and N.~Ayache, \emph{{A Riemannian framework for tensor
  computing}}, \href{https://doi.org/10.1007/s11263-005-3222-z}{\emph{Int. J.
  Computer Vis.} {\bfseries 66} (2006) 41}.

\bibitem{congedo}
A.~Barachant, S.~Bonnet, M.~Congedo and C.~Jutten, \emph{{Multiclass
  Brain–Computer Interface Classification by Riemannian Geometry}},
  \href{https://doi.org/10.1109/TBME.2011.2172210}{\emph{IEEE Trans. Biomed.
  Eng.} {\bfseries 59} (2012) 920}.

\bibitem{moakher2}
M.~Moakher, \emph{{On the Averaging of Symmetric Positive-Definite Tensors}},
  \href{https://doi.org/10.1007/s10659-005-9035-z}{\emph{J. Elasticity}
  {\bfseries 82} (2006) 273}.

\bibitem{arnaudon}
M.~Arnaudon, F.~Barbaresco and L.~Yang, \emph{{Riemannian Medians and Means
  With Applications to Radar Signal Processing}},
  \href{https://doi.org/10.1109/JSTSP.2013.2261798}{\emph{IEEE J. Sel. Topics
  Signal Process.} {\bfseries 7} (2013) 595}.

\bibitem{arnaudon1}
M.~{Arnaudon}, L.~{Yang} and F.~{Barbaresco}, \emph{{Stochastic algorithms for
  computing p-means of probability measures, geometry of radar Toeplitz
  covariance matrices and applications to HR Doppler processing}},  in
  \emph{12th International Radar Symposium (IRS)}, pp.~651--656, 2011.

\bibitem{image1}
S.~{Jayasumana}, R.~{Hartley}, M.~{Salzmann}, H.~{Li} and M.~{Harandi},
  \emph{{Kernel Methods on the Riemannian Manifold of Symmetric Positive
  Definite Matrices}},  in \emph{IEEE Conference on Computer Vision and Pattern
  Recognition}, pp.~73--80, 2013,
  \href{https://doi.org/10.1109/CVPR.2013.17}{DOI}.

\bibitem{image2}
L.~{Zheng}, G.~{Qiu}, J.~{Huang} and J.~{Duan}, \emph{{Fast and accurate
  Nearest Neighbor search in the manifolds of symmetric positive definite
  matrices}},  in \emph{IEEE International Conference on Acoustics, Speech and
  Signal Processing (ICASSP)}, pp.~3804--3808, 2014,
  \href{https://doi.org/10.1109/ICASSP.2014.6854313}{DOI}.

\bibitem{dong}
G.~{Dong} and G.~{Kuang}, \emph{{Target Recognition in SAR Images via
  Classification on Riemannian Manifolds}},
  \href{https://doi.org/10.1109/LGRS.2014.2332076}{\emph{IEEE Geosci. Remote
  Sens. Lett.} {\bfseries 12} (2015) 199}.

\bibitem{tuzel}
O.~{Tuzel}, F.~{Porikli} and P.~{Meer}, \emph{{Pedestrian Detection via
  Classification on Riemannian Manifolds}},
  \href{https://doi.org/10.1109/TPAMI.2008.75}{\emph{IEEE Trans. Pattern Anal.
  Mach. Intell.} {\bfseries 30} (2008) 1713}.

\bibitem{caseiro}
R.~{Caseiro}, J.~F. {Henriques}, P.~{Martins} and J.~{Batista}, \emph{{A
  nonparametric Riemannian framework on tensor field with application to
  foreground segmentation}},  in \emph{International Conference on Computer
  Vision}, pp.~1--8, 2011,
  \href{https://doi.org/10.1109/ICCV.2011.6126218}{DOI}.

\bibitem{cheng2013novel}
G.~Cheng and B.~C. Vemuri, \emph{{A novel dynamic system in the space of {SPD}
  matrices with applications to appearance tracking}},
  \href{https://doi.org/10.1137/110853376}{\emph{SIAM J. Imaging Sci.}
  {\bfseries 6} (2013) 592}.

\bibitem{said2017riemannian}
S.~Said, L.~Bombrun, Y.~Berthoumieu and J.~H. Manton, \emph{{Riemannian
  Gaussian distributions on the space of symmetric positive definite
  matrices}}, \href{https://doi.org/10.1109/TIT.2017.2653803}{\emph{IEEE Trans.
  Information Theory} {\bfseries 63} (2017) 2153}
  [\href{https://arxiv.org/abs/1507.01760}{{\ttfamily 1507.01760}}].

\bibitem{said2017gaussian}
S.~Said, H.~Hajri, L.~Bombrun and B.~C. Vemuri, \emph{{Gaussian distributions
  on Riemannian symmetric spaces: statistical learning with structured
  covariance matrices}},
  \href{https://doi.org/10.1109/TIT.2017.2713829}{\emph{IEEE Trans. Information
  Theory} {\bfseries 64} (2017) 752}
  [\href{https://arxiv.org/abs/1607.06929}{{\ttfamily 1607.06929}}].

\bibitem{said2019warped}
S.~Said, L.~Bombrun and Y.~Berthoumieu, \emph{{Warped Riemannian metrics for
  location-scale models}},  in \emph{Geometric Structures of Information}
  (F.~Nielsen, ed.), pp.~251--296.
\newblock Springer, Switzerland, 2019.
\newblock \href{https://doi.org/10.1007/978-3-030-02520-5}{DOI}.

\bibitem{afsari}
B.~Afsari, \emph{{Riemannian $L^p$ center of mass: Existence, uniqueness, and
  convexity}},
  \href{https://doi.org/10.1090/S0002-9939-2010-10541-5}{\emph{Proc. Amer.
  Math. Soc.} {\bfseries 139} (2011) 655}.

\bibitem{moakher1}
M.~Moakher, \emph{{A differential geometric approach to the geometric mean of
  symmetric positive-definite matrices}},
  \href{https://doi.org/10.1137/S0895479803436937}{\emph{SIAM J. Matrix Anal.
  Appl.} {\bfseries 26} (2005) 735}.

\bibitem{mathieu2019continuous}
E.~Mathieu, C.~Le~Lan, C.~J. Maddison, R.~Tomioka and Y.~Teh, \emph{{Continuous
  hierarchical representations with Poincar\'e variational auto-encoders}},  in
  \emph{Advances in neural information processing systems (NeurIPS2019)}
  (H.~Wallach, H.~Larochelle, A.~Beygelzimer, F.~d'~Alch\'{e}-Buc, E.~Fox and
  R.~Garnett, eds.), vol.~32, pp.~12565--12576, Curran Associates, Inc., 2019,
  [\href{https://arxiv.org/abs/1901.06033}{{\ttfamily 1901.06033}}].

\bibitem{ovinnikov2020poincare}
I.~Ovinnikov, \emph{{Poincar\'e Wasserstein Autoencoder}},  in \emph{Bayesian
  Deep Learning Workshop (NeurIPS 2018)}, 2019,
  [\href{https://arxiv.org/abs/1901.01427}{{\ttfamily 1901.01427}}].

\bibitem{gerald2020node}
T.~Gerald, H.~Zaatiti, H.~Hajri, N.~Baskiotis and O.~Schwander, \emph{{From
  node embedding to community embedding: A hyperbolic approach}},
  [\href{https://arxiv.org/abs/1907.01662}{{\ttfamily 1907.01662}}].

\bibitem{MehtaBook}
M.~L. Mehta, \emph{Random Matrices}, Pure and Applied Mathematics. Elsevier
  Science, 2004.

\bibitem{ForresterBook}
P.~J. Forrester, \emph{Log-gases and random matrices}, vol.~34 of \emph{London
  Mathematical Society Monographs Series}. Princeton University Press,
  Princeton, NJ, 2010,
  \href{https://doi.org/10.1515/9781400835416}{10.1515/9781400835416}.

\bibitem{livan2018introduction}
G.~Livan, M.~Novaes and P.~Vivo, \emph{Introduction to random matrices: theory
  and practice}, vol.~26 of \emph{SpringerBriefs in Mathematical Physics}.
  Springer, Switzerland, 2018,
  \href{https://doi.org/10.1007/978-3-319-70885-0}{10.1007/978-3-319-70885-0},
  [\href{https://arxiv.org/abs/math-ph/0402061}{{\ttfamily math-ph/0402061}}].

\bibitem{muirhead}
R.~J. Muirhead, \emph{{Aspects of multivariate statistical theory}}, vol.~197
  of \emph{Wiley Series in Probability and Statistics}. Wiley, 2009.

\bibitem{AFNvM}
M.~Adler, P.~J. Forrester, T.~Nagao and P.~van Moerbeke, \emph{Classical skew
  orthogonal polynomials and random matrices},
  \href{https://doi.org/10.1023/A:1018644606835}{\emph{J. Stat. Phys.}
  {\bfseries 99} (2000) 141}
  [\href{https://arxiv.org/abs/solv-int/9907001}{{\ttfamily
  solv-int/9907001}}].

\bibitem{Forrester:Vicious}
P.~J. Forrester, \emph{Vicious random walkers in the limit of a large number of
  walkers}, \href{https://doi.org/10.1007/BF01016779}{\emph{J. Stat. Phys.}
  {\bfseries 56} (1989) 767}.

\bibitem{Tierz:02}
M.~Tierz, \emph{{Soft matrix models and Chern-Simons partition functions}},
  \href{https://doi.org/10.1142/S0217732304014100}{\emph{Mod. Phys. Lett.}
  {\bfseries A19} (2004) 1365}
  [\href{https://arxiv.org/abs/hep-th/0212128}{{\ttfamily hep-th/0212128}}].

\bibitem{szeg1939orthogonal}
G.~Szeg\H{o}, \emph{Orthogonal polynomials}, vol.~23. American Mathematical
  Soc., 1939.

\bibitem{schwartzman2006random}
A.~Schwartzman, \emph{{Random ellipsoids and false discovery rates: Statistics
  for diffusion tensor imaging data}}, Ph.D. thesis, Stanford University, 2006.

\bibitem{schwartzman2016lognormal}
A.~Schwartzman, \emph{{Lognormal distributions and geometric averages of
  symmetric positive definite matrices}},
  \href{https://doi.org/10.1111/insr.12113}{\emph{International Stat. Rev.}
  {\bfseries 84} (2016) 456}.

\bibitem{AltlandZirnbauer:96}
A.~Altland and M.~R. Zirnbauer, \emph{{Nonstandard symmetry classes in
  mesoscopic normal-superconducting hybrid structures}},
  \href{https://doi.org/10.1103/PhysRevB.55.1142}{\emph{Phys. Rev. B}
  {\bfseries 55} (1997) 1142}
  [\href{https://arxiv.org/abs/cond-mat/9602137}{{\ttfamily
  cond-mat/9602137}}].

\bibitem{ZirnbauerReview}
M.~R. Zirnbauer, \emph{{Symmetry Classes}},  in \emph{{The Oxford Handbook of
  Random Matrix Theory}} (G.~Akemann, J.~Baik and P.~Di~Francesco, eds.),
  Oxford Handbooks in Mathematics, ch.~3.
\newblock Oxford University Press, Great Clarendon Street, Oxford, UK, 2011.
\newblock [\href{https://arxiv.org/abs/1001.0722}{{\ttfamily 1001.0722}}].

\bibitem{CMReview}
M.~Caselle and U.~Magnea, \emph{{Random matrix theory and symmetric spaces}},
  \href{https://doi.org/10.1016/j.physrep.2003.12.004}{\emph{Phys. Rept.}
  {\bfseries 394} (2004) 41}
  [\href{https://arxiv.org/abs/cond-mat/0304363}{{\ttfamily
  cond-mat/0304363}}].

\bibitem{edelman2020generalized}
A.~Edelman and S.~Jeong, \emph{{The Generalized Cartan Decomposition for
  Classical Random Matrix Ensembles}},
  [\href{https://arxiv.org/abs/2011.08087}{{\ttfamily 2011.08087}}].

\bibitem{Dyson3}
F.~J. Dyson, \emph{{The Threefold Way. Algebraic Structure of Symmetry Groups
  and Ensembles in Quantum Mechanics}},
  \href{https://doi.org/10.1063/1.1703863}{\emph{J. Math. Phys.} {\bfseries 3}
  (1962) 1199}.

\bibitem{romo2012unitary}
M.~Romo and M.~Tierz, \emph{{Unitary Chern-Simons matrix model and the Villain
  lattice action}},
  \href{https://doi.org/10.1103/PhysRevD.86.045027}{\emph{Phys. Rev. D}
  {\bfseries 86} (2012) 045027}
  [\href{https://arxiv.org/abs/1103.2421}{{\ttfamily 1103.2421}}].

\bibitem{giasemidis2014torus}
G.~Giasemidis and M.~Tierz, \emph{{Torus knot polynomials and susy Wilson
  loops}}, \href{https://doi.org/10.1007/s11005-014-0724-z}{\emph{Lett. Math.
  Phys.} {\bfseries 104} (2014) 1535}
  [\href{https://arxiv.org/abs/1401.8171}{{\ttfamily 1401.8171}}].

\bibitem{takahashi2014oscillatory}
Y.~Takahashi and M.~Katori, \emph{{Oscillatory matrix model in Chern-Simons
  theory and Jacobi-theta determinantal point process}},
  \href{https://doi.org/10.1063/1.4894235}{\emph{J. Math. Phys.} {\bfseries 55}
  (2014) 093302} [\href{https://arxiv.org/abs/1312.5848}{{\ttfamily
  1312.5848}}].

\bibitem{garcia2020matrix}
D.~Garcia-Garcia and M.~Tierz, \emph{{Matrix models for classical groups and
  Toeplitz$\pm$Hankel minors with applications to Chern-Simons theory and
  fermionic models}}, \href{https://doi.org/10.1088/1751-8121/ab9b4d}{\emph{J.
  Phys. A} {\bfseries 53} (2020) }
  [\href{https://arxiv.org/abs/1901.08922}{{\ttfamily 1901.08922}}].

\bibitem{Andreief}
C.~Andr\'{e}ief, \emph{{Note sur une relation entre les int\'{e}grales
  d\'{e}finies des produits des fonctions}}, {\emph{M\'{e}m . Soc. Sci. Phys.
  Nat. Bordeaux} {\bfseries 2} (1886) 1}.

\bibitem{ForresterMeet}
P.~J. Forrester, \emph{{Meet Andr\'{e}ief, Bordeaux 1886, and Andreev, Kharkov
  1882–-1883}}, \href{https://doi.org/10.1142/S2010326319300018}{\emph{Random
  Matrices: Theory and Applications} {\bfseries 08} (2019) 1930001}
  [\href{https://arxiv.org/abs/1806.10411}{{\ttfamily 1806.10411}}].

\bibitem{GST}
G.~Giasemidis, R.~J. Szabo and M.~Tierz, \emph{{Supersymmetric gauge theories,
  Coulomb gases and Chern-Simons matrix models}},
  \href{https://doi.org/10.1103/PhysRevD.89.025016}{\emph{Phys. Rev. D}
  {\bfseries 89} (2014) 025016}
  [\href{https://arxiv.org/abs/1310.3122}{{\ttfamily 1310.3122}}].

\bibitem{Baxter:63}
R.~J. Baxter, \emph{Statistical mechanics of a one-dimensional coulomb system
  with a uniform charge background},
  \href{https://doi.org/10.1017/S0305004100003790}{\emph{Math. Proc. Cambridge
  Phil. Soc.} {\bfseries 59} (1963) 779–787}.

\bibitem{Dhar:2017grt}
A.~Dhar, A.~Kundu, S.~N. Majumdar, S.~Sabhapandit and G.~Schehr, \emph{{Exact
  Extremal Statistics in the Classical 1D Coulomb Gas}},
  \href{https://doi.org/10.1103/PhysRevLett.119.060601}{\emph{Phys. Rev. Lett.}
  {\bfseries 119} (2017) 060601}
  [\href{https://arxiv.org/abs/1704.08973}{{\ttfamily 1704.08973}}].

\bibitem{Dhar:2018}
A.~Dhar, A.~Kundu, S.~N. Majumdar, S.~Sabhapandit and G.~Schehr, \emph{{Extreme
  statistics and index distribution in the classical 1d Coulomb gas}},
  \href{https://doi.org/10.1088/1751-8121/aac75f}{\emph{J. Phys. A} {\bfseries
  51} (2018) 295001} [\href{https://arxiv.org/abs/1802.10374}{{\ttfamily
  1802.10374}}].

\bibitem{Aganagic:2002wv}
M.~Aganagic, A.~Klemm, M.~Marino and C.~Vafa, \emph{{Matrix model as a mirror
  of Chern-Simons theory}},
  \href{https://doi.org/10.1088/1126-6708/2004/02/010}{\emph{JHEP} {\bfseries
  02} (2004) 010} [\href{https://arxiv.org/abs/hep-th/0211098}{{\ttfamily
  hep-th/0211098}}].

\bibitem{Marino:2004eq}
M.~Marino, \emph{{Les Houches lectures on matrix models and topological
  strings}},  10, 2004, [\href{https://arxiv.org/abs/hep-th/0410165}{{\ttfamily
  hep-th/0410165}}].

\bibitem{deBruijn}
N.~G. de~Bruijn, \emph{{On some multiple integrals involving determinants}},
  {\emph{J. Indian Math. Soc.} {\bfseries 19} (1955) 133}.

\bibitem{Wimmer}
M.~Wimmer, \emph{Efficient numerical computation of the pfaffian for dense and
  banded skew-symmetric matrices},
  \href{https://doi.org/10.1145/2331130.2331138}{\emph{ACM Trans. Math.
  Software} {\bfseries 38} (2012) 1–}
  [\href{https://arxiv.org/abs/1102.3440}{{\ttfamily 1102.3440}}].

\bibitem{AHvM:Pfaff}
M.~Adler, E.~Horozov and P.~van Moerbeke, \emph{{The Pfaff lattice and
  skew-orthogonal polynomials}},
  \href{https://doi.org/10.1155/S107379289900029X}{\emph{Int. Math. Res. Not.}
  {\bfseries 11} (1999) 569}
  [\href{https://arxiv.org/abs/solv-int/9903005}{{\ttfamily
  solv-int/9903005}}].

\bibitem{Chang:Pfaff}
X.-K. Chang, Y.~He, X.-B. Hu and S.-H. Li, \emph{Partial-skew-orthogonal
  polynomials and related integrable lattices with pfaffian tau-functions},
  \href{https://doi.org/10.1007/s00220-018-3273-y}{\emph{Commun. Math. Phys.}
  {\bfseries 364} (2018) 1069–1119}
  [\href{https://arxiv.org/abs/1712.06382}{{\ttfamily 1712.06382}}].

\bibitem{said2017riemannianb}
S.~Said, N.~Le~Bihan and J.~H. Manton, \emph{Riemannian gaussian distributions
  on the space of positive-definite quaternion matrices},  in
  \emph{International Conference on Geometric Science of Information} (B.~F.
  Nielsen~F., ed.), vol.~10589 of \emph{Lecture Notes in Computer Science},
  pp.~709--716, Springer, 2017,
  [\href{https://arxiv.org/abs/1703.09940}{{\ttfamily 1703.09940}}],
  \href{https://doi.org/https://doi.org/10.1007/978-3-319-68445-1_82}{DOI}.

\bibitem{fisher1984walks}
M.~E. Fisher, \emph{Walks, walls, wetting, and melting},
  \href{https://doi.org/10.1007/BF01009436}{\emph{J. Stat. Phys.} {\bfseries
  34} (1984) 667}.

\bibitem{karlin1959coincidence}
S.~Karlin and J.~McGregor, \emph{Coincidence probabilities},
  \href{https://doi.org/10.2140/pjm.1959.9.1141}{\emph{Pacific J. Math.}
  {\bfseries 9} (1959) 1141}.

\bibitem{grabiner1999brownian}
D.~J. Grabiner, \emph{{Brownian motion in a Weyl chamber, non-colliding
  particles, and random matrices}},
  \href{https://doi.org/10.1016/S0246-0203(99)80010-7}{\emph{Annales de l'IHP
  B: Probabilit{\'e}s et statistiques} {\bfseries 35} (1999) 177}
  [\href{https://arxiv.org/abs/math/9708207}{{\ttfamily math/9708207}}].

\bibitem{katori2016bessel}
M.~Katori, \emph{{Bessel processes, Schramm-Loewner evolution, and the Dyson
  model}}, vol.~11 of \emph{SpringerBriefs in Mathematical Physics}. Springer
  Singapore, 2016,
  \href{https://doi.org/10.1007/978-981-10-0275-5}{10.1007/978-981-10-0275-5}.

\bibitem{BaikDeiftSuidanBook}
J.~Baik, P.~Deift and T.~Suidan, \emph{Combinatorics and random matrix theory},
  vol.~172 of \emph{Graduate Studies in Mathematics}. American Mathematical
  Society, Providence, RI, 2016.

\bibitem{dyson1962brownian}
F.~J. Dyson, \emph{A brownian-motion model for the eigenvalues of a random
  matrix}, \href{https://doi.org/10.1063/1.1703862}{\emph{J. Math. Phys.}
  {\bfseries 3} (1962) 1191}.

\bibitem{coifman2006diffusion}
R.~R. Coifman and S.~Lafon, \emph{Diffusion maps},
  \href{https://doi.org/10.1016/j.acha.2006.04.006}{\emph{Applied and
  computational harmonic analysis} {\bfseries 21} (2006) 5}.

\bibitem{botev2010kernel}
Z.~I. Botev, J.~F. Grotowski and D.~P. Kroese, \emph{Kernel density estimation
  via diffusion}, \href{https://doi.org/10.1214/10-aos799}{\emph{Ann. Statist.}
  {\bfseries 38} (2010) 2916}
  [\href{https://arxiv.org/abs/1011.2602}{{\ttfamily 1011.2602}}].

\bibitem{deHaroTierz:04}
S.~de~Haro and M.~Tierz, \emph{{Brownian motion, Chern-Simons theory, and 2-D
  Yang-Mills}},
  \href{https://doi.org/10.1016/j.physletb.2004.09.033}{\emph{Phys. Lett. B}
  {\bfseries 601} (2004) 201}
  [\href{https://arxiv.org/abs/hep-th/0406093}{{\ttfamily hep-th/0406093}}].

\bibitem{TierzDolivet}
Y.~Dolivet and M.~Tierz, \emph{{Chern-Simons matrix models and Stieltjes-Wigert
  polynomials}}, \href{https://doi.org/10.1063/1.2436734}{\emph{J. Math. Phys.}
  {\bfseries 48} (2007) 023507}
  [\href{https://arxiv.org/abs/hep-th/0609167}{{\ttfamily hep-th/0609167}}].

\bibitem{tierz2010schur}
M.~Tierz, \emph{Schur polynomials and biorthogonal random matrix ensembles},
  \href{https://doi.org/10.1063/1.3377965}{\emph{J. Math. Phys.} {\bfseries 51}
  (2010) 063509}.

\bibitem{baik2007random}
J.~Baik and T.~M. Suidan, \emph{Random matrix central limit theorems for
  nonintersecting random walks},
  \href{https://doi.org/10.1214/009117906000001105}{\emph{Ann. Probab.}
  {\bfseries 35} (2007) 1807}
  [\href{https://arxiv.org/abs/math/0605212}{{\ttfamily math/0605212}}].

\bibitem{katori2002scaling}
M.~Katori and H.~Tanemura, \emph{Scaling limit of vicious walks and two-matrix
  model}, \href{https://doi.org/10.1103/physreve.66.011105}{\emph{Phys. Rev. E}
  {\bfseries 66} (2002) 011105}
  [\href{https://arxiv.org/abs/cond-mat/0203549}{{\ttfamily
  cond-mat/0203549}}].

\bibitem{katori2004symmetry}
M.~Katori and H.~Tanemura, \emph{Symmetry of matrix-valued stochastic processes
  and noncolliding diffusion particle systems},
  \href{https://doi.org/10.1063/1.1765215}{\emph{J. Math. Phys.} {\bfseries 45}
  (2004) 3058} [\href{https://arxiv.org/abs/1712.07903}{{\ttfamily
  1712.07903}}].

\bibitem{forrester2019orthogonal}
P.~J. Forrester, J.~R. Ipsen, D.-Z. Liu and L.~Zhang, \emph{{Orthogonal and
  symplectic Harish-Chandra integrals and matrix product ensembles}},
  \href{https://doi.org/10.1142/S2010326319500151}{\emph{Random Matrices:
  Theory and Applications} {\bfseries 8} (2019) 1950015}
  [\href{https://arxiv.org/abs/1711.10691}{{\ttfamily 1711.10691}}].

\bibitem{hendriks1990nonparametric}
H.~Hendriks, \emph{{Nonparametric estimation of a probability density on a
  Riemannian manifold using Fourier expansions}},
  \href{https://doi.org/10.1214/aos/1176347628}{\emph{Ann. Statist.} {\bfseries
  18} (1990) 832}.

\bibitem{walter1977properties}
G.~G. Walter, \emph{{Properties of Hermite series estimation of probability
  density}}, \href{https://doi.org/10.1214/aos/1176344013}{\emph{Ann. Statist.}
  {\bfseries 5} (1977) 1258}.

\bibitem{Forrester:crystal}
P.~J. Forrester, \emph{{Properties of an exact crystalline many-body ground
  state}}, \href{https://doi.org/10.1007/BF02188665}{\emph{J. Stat. Phys.}
  {\bfseries 76} (1994) 331}.

\bibitem{Forrester:new}
P.~J. Forrester, \emph{{Global and local scaling limits for the $\beta = 2$
  Stieltjes-Wigert random matrix ensemble}},
  [\href{https://arxiv.org/abs/2011.11783}{{\ttfamily 2011.11783}}].

\bibitem{mays2020tracy}
A.~Mays, A.~Ponsaing and G.~Schehr, \emph{{Tracy-Widom distributions for the
  Gaussian orthogonal and symplectic ensembles revisited: a skew-orthogonal
  polynomials approach}},  [\href{https://arxiv.org/abs/2007.14597}{{\ttfamily
  2007.14597}}].

\end{thebibliography}\endgroup
\end{document}